\documentclass[preprint]{elsarticle}  
\usepackage{graphicx}
\usepackage{color}
\usepackage[normalem]{ulem}
\begin{document}

\title{Factorization in large-scale many-body calculations\tnoteref{t1}}
\tnotetext[t1]{UCRL number: LLNL-JRNL-624065}
\author[sdsu]{Calvin W. Johnson}
\ead{cjohnson@mail.sdsu.edu}
\address[sdsu]{Department of Physics, San Diego State University,
5500 Campanile Drive, San Diego, CA 92182-1233}
\author[llnl]{W. Erich Ormand}
\ead{ormand1@llnl.gov}
\address[llnl]{Lawrence Livermore National  Laboratory, P.O. Box 808, L-414,
Livermore, CA 94551}
\author[sdsu,llnl,harvard]{Plamen G. Krastev}
\ead{plamenkrastev@fas.harvard.edu}
\address[harvard]{Harvard University, Research Computing, Faculty of Arts and Sciences,
38 Oxford Street, Cambridge, MA 02138}

\begin{abstract}
One  approach for solving interacting many-fermion systems  is the configuration-interaction
method, also sometimes called the interacting shell model, where one
finds eigenvalues of the Hamiltonian in a many-body basis of Slater
determinants (antisymmetrized products of single-particle
wavefunctions).  The resulting Hamiltonian matrix is typically very
sparse, but for large systems the nonzero matrix elements
can nonetheless require terabytes or more of storage. An alternate algorithm, 
applicable to a broad class of systems with symmetry, in our case rotational
invariance, is to exactly factorize
both the basis and the interaction using additive/multiplicative
quantum numbers; such an algorithm recreates the many-body matrix 
elements on the fly and can reduce the storage
requirements by an order of magnitude or more. We discuss
factorization in general and introduce a novel, generalized factorization method, 
essentially a `double-factorization' which speeds up basis generation and set-up of required arrays. 
Although we emphasize techniques, we also place 
factorization in the context of a specific (unpublished) 
configuration-interaction code, BIGSTICK, which runs both on serial
and parallel machines, and discuss the savings in memory due to factorization.
\end{abstract}

\maketitle

\section{Introduction}

The quantum mechanics of many-body systems, specifically when the
number of particles is between three and a few hundred, is a
theoretical and computational challenge. Often these systems have
exact symmetries  that can be either a curse or a
gift. The quantum numbers associated with each symmetry should be treated exactly, hindering
approximations. Conversely, some quantum numbers can \textit{aid}
calculation by excluding trivial matrix elements via selection rules. The algorithmic
exploitation of quantum numbers for efficient calculations in
many-body systems is the theme of this paper.

Central to the methods described here are abelian symmetries: in
practical terms this means the quantum numbers for many-body systems
are simply the sum or product of the quantum numbers of the
constituent single-particle states. Examples include parity and,
most importantly for us, the $z$-component of angular momentum.  
We focus on systems of fermions with rotationally
invariant Hamiltonians: multi-electron atoms, atomic nuclei, and
general fermions (e.g., cold atoms) in a spherically symmetric trap.
Also key  is  the presence of two `species' of
fermions, e.g. protons and neutrons, spin-up and spin-down
electrons, etc.  While there is a long menu of methods and
approximations to tackle such systems, we consider only 
configuration-interaction (CI) calculations, finding low-lying
eigenvalues of the Hamiltonian matrix computed in a large-dimension
 basis \cite{Sh98,BG77,br88,ca05,Cook98,Je07,Lo55CI,We61CI,SS99,ORJJ88,OJS90}.
 CI uses a many-body basis of Slater determinants,
antisymmetrized products of single-particle wavefunctions that can
be rendered trivially orthonormal and, as we will see, which lend to
rapid computation of Hamiltonian matrix elements. 

The advantages of
CI are (a) it is fully microscopic; (b) allows for arbitrary
single-particle basis; (c) allows for arbitrary form of the two-body
interaction, i.e., no restriction on locality or momentum
dependence, etc.; (d) is equally effective for even or odd numbers
of particles and has no difficulty with `open shell' systems and (e)
can compute multiple excited states  with the same quantum numbers.
The main disadvantage of CI is it is not size-extensive and so contains
a large number of unlinked diagrams that must be canceled \cite{Sh98}, leading to a
slow convergence as the dimension of the model space increases. Hence
CI bases often have very large dimensions, and one often turns to an effective interaction that partially sums over many
single-particle states.  Other many-body methods have different sets of advantages and
disadvantages, of course.

Configuration-interaction calculations expand the many-body
wavefunction in an orthonormal basis,
\begin{equation}
| \Psi \rangle = \sum_\alpha c_\alpha | \alpha \rangle. \label{CIexpand}
\end{equation}
Subsequently the central computational goal of a CI calculation is to find
eigenvalues and eigenvectors of a very large matrix $H_{\alpha \beta} =
\langle \alpha | \hat{H} |  \beta \rangle$, that is, to
solve
\begin{equation}
\mathbf{H} \vec{c}_n= E_n \vec{c}_n.
\end{equation}
 Because one is generally interested only in
low-lying states, typically the lowest 5-20 states, one can use
Arnoldi methods such as the Lanczos algorithm \cite{GVL, Lanczos},
where one iteratively transforms the Hamiltonian to tridiagonal form:
\begin{equation}
\mathbf{H}|v_n\rangle = b_{n-1} |v_{n-1}\rangle + a_n |v_n\rangle
+ b_n |v_{n+1}\rangle.
\end{equation}
This creates a unitary transformation to a new basis.  The advantage of
Lanczos over other unitary transformations to tridiagonal form such as Householder \cite{GVL}
is that one does not need to complete the transformation. If we truncate after
$n-1$ Lanczos iterations, leaving us with the approximate matrix
\begin{equation}
\left (
\begin{array}{ccccccc}
a_1  & b_1 & 0 & 0 & \ldots & 0 & 0 \\
b_1  & a_2 & b_2  & 0 & \ldots & 0 & 0 \\
0 & b_2 & a_3 & b_3   & \ldots & 0 & 0 \\
0 & 0 & b_3 & a_4   & \ldots & 0 & 0 \\
\vdots & & & & & & \vdots \\
0 & 0 & \ldots & & & a_{n-1} & b_n \\
0 & 0 & \ldots & & & b_{n} & a_n
\end{array}
\right ),
\end{equation}
the extremal eigenvalues converge quickly \cite{Lanczos}, which one
can understand through the lens of the classical moments problem
\cite{LanczosMoment}. The downside of Lanczos is that, due to
numerical round-off error the Lanczos vectors $|v_n \rangle$ lose
orthogonality and must be forcibly orthonormalized, which is why
Householder is often preferred when one must \textit{completely}
transform a matrix to tridiagonal form.

The important point to take away is that matrix-vector
multiplication is the fundamental operation. In large cases, in the
so-called $M$-scheme (described below), the dimensions can be 
upwards of $10^{7-10}$; the associated many-body Hamiltonian matrix is very sparse,
with only one matrix element out of ten thousand or a million nonzero, if the 
embedded interaction is two-body in nature. Specific
examples for nuclei are given in Table I, demonstrating that storage of
just nonzero matrix elements requires hundreds of gigabytes,
terabytes, and even petabytes; if one uses 3-body interactions,
discussed in Section \ref{3body} and illustrated in Table
\ref{threebody}, the matrices are significantly less sparse and the
storage demands even higher. In fact, one can argue that it is not
the dimensionality of the CI vector as in Eq.~(\ref{CIexpand}) but
the number of nonzero matrix elements that governs the computational
difficulty of a CI calculation--a completely dense matrix of
dimension $10^6$ has much higher demand on memory than a matrix of dimension $10^8$
but with sparsity $10^{-6}$.

\begin{table}
\caption{Some model spaces for atomic nuclei and their  $M$-scheme ($M=0$) dimensions and
the sparsity for a two-body Hamiltonian. ``Storage'' refers to memory
requirements in gigabytes for the
nonzero matrix elements.
The model spaces are described in detail in Appendix B; $N_\mathrm{shell}$ includes
all configurations, while $N_\mathrm{max}$ is a truncation on the non-interacting
energy.
\label{nuclear_sparsity}}
\begin{tabular}{|c|c|c|c|c|}
\hline
Nuclide  &  space  & basis  & sparsity & storage  \\
         &         &  dim.  &          &  (GB) \\
\hline
$^{28}$Si  & $sd$ & $9.4 \times 10^4$ & $6 \times 10^{-3}$  &  0.2\\
$^{52}$Fe  & $pf$ & $1.1 \times 10^8$ & $1 \times 10^{-5}$ & 720 \\
$^{56}$Ni  & $pf$ & $1.1 \times 10^9$ & $2 \times 10^{-6}$ & 9600\\
$^{4}$He   & $N_\mathrm{shell} =8$ & $2.9 \times 10^{7}$ & $4 \times 10^{-4}$ & 1440 \\
$^{4}$He   & $N_\mathrm{shell} =10$ & $2.7 \times 10^{8}$ & $1 \times 10^{-4}$ & 36,000 \\
$^{4}$He   & $N_\mathrm{max} =16$ & $6.3 \times 10^{6}$ & $1 \times 10^{-3}$ & 200 \\
$^{4}$He   & $N_\mathrm{max} =22$ & $8.6 \times 10^{7}$ & $3 \times 10^{-4}$ & 9600 \\
$^{12}$C   & $N_\mathrm{shell} =3$ & $8.2 \times 10^{7}$ & $1 \times 10^{-5}$ & 400 \\
$^{12}$C   & $N_\mathrm{shell} =4$ & $5.9 \times 10^{11}$ & $8 \times 10^{-9}$ & $1
\times 10^7$ \\
$^{12}$C   & $N_\mathrm{max} =8$ & $5.9 \times 10^{8}$ & $4 \times 10^{-6}$ & 5200 \\
$^{13}$C   & $N_\mathrm{max} =6$ & $3.8 \times 10^{7}$ & $4 \times 10^{-5}$ & 210 \\

\hline
\end{tabular}
\end{table}

There are two approaches to the problem of a large, sparse matrix.
First, one can simply store the nonzero matrix elements. In nuclear
physics the CI code of Whitehead \textit{et al.} \cite{Lanczos} and later
the OXBASH code \cite{OXBASH} and its successor NuShell \cite{NuShell}
stored the matrix elements on disk, but for modern computers reading
data from disk dramatically slows down the algorithm. Alternately,
one can store the matrix elements in RAM, which is much faster but
for all but the most modest of problems requires distribution across
hundreds or thousands of nodes on a parallel computer, as done with
the MFDn code \cite{MFD}.

Not only does storage of the many-body Hamiltonian matrix elements
put an enormous drain on memory resources, it is wasteful:
the nonzero matrix elements are not unique but have an enormous redundancy,
each one reused many times over, as  matrix elements between pairs of
particles are generally
unique, but in a many-body system there are a large number of
possible combinations of inert spectators.

An alternate to storage of the many-body matrix elements is to
recreate them on the fly, which by reducing redundancy requires one or two
orders of magnitude less memory.
On-the-fly recalculation can be made surprisingly efficient by
factorizing the problem into complementary parts, using quantum
numbers.  (Quantum numbers, which
label irreducible representations of symmetry groups, are generally
found as eigenvalues of commuting operators; for a review see 
 \ref{quantum_numbers}.  Note that although we focus on continuous
symmetries, that is rotation, factorization will work for any abelian symmetry,
for which one can compute the quantum number for a many-body state by simply
adding or multiplying the quantum numbers of the single-particle states. Therefore
the methods described herein could be carried over to discrete symmetries, not
only parity but also point-group symmetries as long as there is an abelian
subgroup such as the cyclic group of order $n$, $C_n$.)  In
the next section, we describe how one can, for a broad class of
cases, efficiently factorize the many-body basis, and, in the
following section, we discuss factorization in applying the
Hamiltonian matrix. These factorization methods were used in
several major CI codes, ANTOINE \cite{ANTOINE}, NATHAN\cite{NATHAN}, EICODE \cite{EICODE,Toi06},
NuShellX \cite{NuShell}, 
and our own unpublished codes REDSTICK and BIGSTICK.
Methods similar to factorization have been used in quantum chemistry (see Ref.~\cite{ORJJ88,OJS90}
and references therein), which allowed quantum chemists to reach dimensions of over a billion (but again:
the computational difficulty is not only the dimensionality of the vectors but also the number of
nonzero matrix elements). Factorization has also been used in nuclear structure physics as a 
gateway to approximation schemes \cite{AP01,PD03,PJD04,PD05}.

While the basic ideas are outlined in Ref.~\cite{ANTOINE,NATHAN,Toi06}, in the following
two sections we
present in detail how factorization work, both for the basis and
for the Hamiltonian, and show explicitly how it reduces
memory load. In Section \ref{haiku}, we give a new application that
further exploits this approach and forms the heart of our latest, but unpublished CI code,
 BIGSTICK. Note that we are not yet publishing BIGSTICK, although we hope to make 
it available soon; this paper presents techniques and not a code.  Finally, we discuss
parallelization of the algorithm and its performance.  Most of our examples
are taken from atomic nuclei, but in  \ref{atomic_examples} we also give similar numbers
for the electronic structure of atoms.

All our examples in this paper refer to many-fermion systems.  Because factorization 
algorithms rely upon conserved quantum numbers, they could be applied to many-boson 
systems as well, and we see no particular difficulties in doing so. We do not discuss 
boson systems, however, in this paper.

A note about conventions. We use lower-case Greek letters, typically $\alpha, \beta$
to label  \textit{many-body} basis states,
which as explained below will be Slater determinants.
When we restrict a Slater determinant
to a single species of particle (e.g., protons or neutrons, or spin-up
or spin-down electrons), we typically use $\mu, \nu$ and add a suffix, e.g.  $\mu_p$ or
$\nu_\downarrow$; for a generic, unspecified species we use $x$ and $y$.
\textit{Single-particle} states we label with $i,j,k,l$. Quantum numbers
 associated with single-particle states are denoted by lower-case
letters, e.g. $j$, $m$, $\pi$, while for many-body states we use capital letters
$J$, $M$, and $\Pi$.

\section{Factorization of the basis}

The concept of factorization can be seen most easily in an efficient
representation of the many-body basis. In order to exploit
factorization we must have two (or more) species of fermions,  for
example, protons and neutrons, or spin-up and spin-down electrons (as done sometimes
in quantum chemistry \cite{ORJJ88,OJS90}),
or two spin-species of cold atoms.  For generality we label these species
$x$ and $y$. Then any wavefunction can be expanded in a sum of
product wavefunctions,
\begin{equation}
| \Psi \rangle = \sum_{\mu \nu} c_{\mu \nu} |\mu_x \rangle | \nu_y \rangle
\label{basis_factored}
\end{equation}
where we already see the basis states factorized.

Like  good physicists, we build our many-body state $| \mu_x \rangle, | \nu_y \rangle $ from simple components, and
start with a finite set of
orthonormal \textit{single-particle} states $\{ \phi_i \}$.
For our purposes, the single-particle states must have as good
quantum numbers total angular momentum, $j$, and $z$-component of
angular momentum, $m$. While for the nuclear case these are assumed
to take on half-integer values, the algorithm can be trivially
generalized to integer values, for example, if one has spin-up and
spin-down electrons as separate species, as described in \ref{atomic_examples}. One generally also assumes
good parity $\pi$.

Thus, one can imagine the single-particle states as eigenstates of a
rotationally invariant single-particle Hamiltonian. The Hamiltonian
that generates these single-particle states is fictitious and is
chosen for convenience; in the limit of an infinite space, the final result
will not depend on the single-particle states. For finite spaces, of course,
the choice of single-particle states can critically affect convergence, and in
nuclear physics one often chooses
harmonic oscillator or mean-field single-particle states. The radial
dependence of each $\phi_i$ only enters into the numerical
values of the interaction matrix elements, which are evaluated
externally and read in as a file.
All we \textit{need} to know for each state $\phi_i$ are the quantum
numbers $j_i, m_i$ and optionally $\pi_i$.  We assume for a given
$j_i$ all possible $m_i$ are allowed.

To illustrate, consider a specific example from the structure of
atomic nuclei.  The $sd$ valence space contains
the $1s_{1/2}$, $0d_{3/2}$ and $0d_{5/2}$ orbits the quantum
numbers of which are given in Table \ref{sdstates}.  Throughout the
rest of this paper we will give examples built  upon this single particle space.
An alternate set of examples from the electronic structure
of atoms can be found in  \ref{atomic_examples}.

\begin{table}
\caption{Ordered list of single-particle states in the $sd$-shell for atomic nuclei.
\label{sdstates}}
\begin{tabular}{|c|c|c|r|}
\hline
State & $l$ & $j$ & $m_j$ \\
\hline
1 & 0 & 1/2 & -1/2 \\
2 & 0 & 1/2 & +1/2 \\
3 & 2 & 3/2 & -3/2 \\
4 & 2 & 3/2 & -1/2 \\
5 & 2 & 3/2 & +1/2 \\
6 & 2 & 3/2 & +3/2 \\
7 & 2 & 5/2 & -5/2 \\
8 & 2 & 5/2 & -3/2 \\
9 & 2 & 5/2 & -1/2 \\
10 & 2 & 5/2 & +1/2 \\
11 & 2 & 5/2 & +3/2 \\
12 & 2 & 5/2 & +5/2 \\
\hline
\end{tabular}
\end{table}

Now that we have the single-particle states, we  construct
the many-body states, also known as Slater determinants, 
using antisymmetrized products of single-particle states.  
A coordinate-space Slater determinant
for $n$ particles is written as
\begin{equation}
\Psi_\mu(\vec{r}_1, \vec{r}_2, \ldots, \vec{r}_n) =
\frac{1}{\sqrt{n!}} \left |
\begin{array}{cccc}
\phi_1(\vec{r}_1) & \phi_2(\vec{r}_1) & \ldots & \phi_n(\vec{r}_1) \\
\phi_1(\vec{r}_2) & \phi_2(\vec{r}_2) & \ldots & \phi_n(\vec{r}_2) \\
\vdots & & \ddots & \\
\phi_1(\vec{r}_n) & \phi_2(\vec{r}_n) & \ldots & \phi_n(\vec{r}_n)
\end{array} \right |
\end{equation}
A Slater
determinant can, however, be compactly represented using second quantization \cite{BG77,LM85}.
Let $\hat{a}_i^\dagger$ be the creation operator associated with
the $i$th state $\phi_i(\vec{r})$; then the occupation- or number-space representation of a
 Slater determinant of $n$ particles is given by
\begin{equation}
| \mu  \rangle = \hat{a}^\dagger_1 \hat{a}^\dagger_2 \ldots
\hat{a}^\dagger_n | 0 \rangle
\end{equation}
where $| 0 \rangle$ is the vacuum state (or an inert/frozen core).
Different Slater determinants will have different combinations of
$\{ \phi_i \}$ and thus use different combinations of creation operator $\{ \hat{a}_i \}$.
So, for example, drawing up the single-particle states
in Table \ref{sdstates}, some possible five-particle states might be
$\hat{a}^\dagger_1 \hat{a}^\dagger_2 \hat{a}^\dagger_3
\hat{a}^\dagger_4 \hat{a}^\dagger_5 | 0 \rangle$,
$\hat{a}^\dagger_1 \hat{a}^\dagger_2 \hat{a}^\dagger_3
\hat{a}^\dagger_4 \hat{a}^\dagger_6 | 0 \rangle$,
$ \hat{a}^\dagger_2 \hat{a}^\dagger_3
\hat{a}^\dagger_{10} \hat{a}^\dagger_{11}  \hat{a}^\dagger_{12} | 0 \rangle$,
and so on.  The ordering is important but only insofar as a different
ordering can lead to a phase due to antisymmetry: $\hat{a}_1^\dagger \hat{a}_2^\dagger
= - \hat{a}_2^\dagger \hat{a}_1^\dagger$, etc.; while we will not emphasize the phase,
getting the phase right is critical.

The Pauli exclusion principle means each single-particle state can
be occupied by at most one particle. This is particularly convenient
for computers, as a bit-representation of the occupation of
single-particle states is natural \cite{Lanczos}. Thus our example
Slater determinants become, respectively, 111110000000,
111101000000, and 011000000111.  Again, the ordering is arbitrary
but must be fixed consistently in order to determine the phase
(often one starts from the right rather than from the left as we
have--the only important thing is consistency keeping track of the
ordering). The sets of bits can be simply stored as integers, and
manipulation of the individual bits is straightforward if somewhat
tedious \cite{Lanczos}. An additional convenience is that if the $\{
\phi_i \}$ form an orthonormal set, the resulting Slater
determinants will also be orthonormal. For 5 protons in the $sd$
valence space (Table \ref{sdstates}), there are a total of $\left (
\begin{array}{c} 12 \\ 5 \end{array} \right) = 792$ possible Slater
determinants.

Not every combination is needed, however, as explained in the next two sections.

\subsection{The $M$-scheme basis}

We invoke the critical assumption that the many-body Hamiltonian is
rotationally invariant, so that both total $\hat{J}^2$ and  total
$\hat{J}_z$ commute with the Hamiltonian and the eigenstates of the
Hamiltonian have both $J$ and $M$ as good quantum numbers,
respectively. This in turn means that the Hamiltonian will not
connect many-body states with different $M$. Therefore we choose an
\textit{M-scheme basis}, that is the many-body basis states all have
the same $M$.  The original Whitehead code \cite{Lanczos}, ANTOINE
\cite{ANTOINE}, and MFDn \cite{MFD} are all $M$-scheme codes. This
is convenient because $m$ is an additive quantum number: to
determine the $M$ for a Slater determinant, one just has to add the
$m_i$ of the occupied single-particle states.

One can  create a \textit{J-scheme basis}, where the basis states
also have fixed $J$; the many-body Hamiltonian is block-diagonal in
$J$. In atomic physics, these are called configuration-state
functions (CSFs).  But this is less straightforward, as each basis
state must be represented as a linear combination of $M$-scheme
basis states. The $J$-scheme basis has both advantages and
disadvantages, which we will not discuss here. Examples of
$J$-scheme codes in nuclear physics include OXBASH \cite{OXBASH} and
its successors NuShell and NuShellX \cite{NuShell},  NATHAN
\cite{NATHAN}, and EICODE \cite{EICODE,Toi06}; OXBASH and NuShell store the many-body Hamiltonian
matrix elements on disk, while the latter three utilize factorization,
but a discussion of factorization with a \textit{J}-scheme basis is
not the intent of this paper.

Using the single-particle states in Table \ref{sdstates},
the state $\hat{a}^\dagger_1 \hat{a}^\dagger_2
\hat{a}^\dagger_3 \hat{a}^\dagger_4 \hat{a}^\dagger_5 | 0 \rangle$
has $M = -3/2$, the state $\hat{a}^\dagger_1 \hat{a}^\dagger_2
\hat{a}^\dagger_3 \hat{a}^\dagger_4 \hat{a}^\dagger_6 | 0 \rangle$
has $M = -1/2$, and the state $ \hat{a}^\dagger_2 \hat{a}^\dagger_3
\hat{a}^\dagger_{10} \hat{a}^\dagger_{11}  \hat{a}^\dagger_{12} | 0
\rangle$ has $M = +7/2$.  While the total number of five-particle
states in this valence space is 792, there are 119 states with $M =
1/2$, 104 with $M =  3/2$, 80 with $M = 5/2$, and 51 with $M = 7/2$,
28 with $M=9/2$, 11 with $M=11/2$, and 3 with $M = 13/2$ ( the same
number apply for $M = -1/2, -3/2$, etc.) For 6 neutrons in the $sd$
valence space, there are a total of $\left ( \begin{array}{c} 12 \\
6 \end{array} \right) = 924$ possible many-body states, but if we
restrict ourselves to $M =0$ there are only 142 states.

As we assume rotational invariance, eigenstates should have good $J$ and $M$,
and the eigenvalue can only depend upon $J$ and not $M$.  The $M$-scheme
eliminates the rotational degeneracy and reduces the size of the basis.

These simple examples are for a single species of particles. With
two species  of particles, the many-body basis becomes more complex,
but factorization allows a compact representation, as we discuss
next.

\subsection{Factorization of the $M$-scheme basis and basis `sectors'}
\label{factor_mscheme}

In the expansion defined by Eq.~(\ref{basis_factored}), both the $x$-species
state $| \mu_x \rangle$ and the $y$-state $| \nu_y \rangle$ are
represented by Slater determinants. Now we can begin to see the
usefulness of factorization. One could represent each final basis
state by a single Slater determinant, by simply combining 
bit-strings, but this is inefficient, because in general any given
$x$-species state $| \mu_x \rangle$ can be combined with more than
one $y$-state $| \nu_y \rangle$ in constructing the basis. This
means one can construct the many-body basis from a small number of
components. We will give more detailed examples below, but let us
consider the case of the $^{27}$Al nucleus, using the $sd$ valence
space. This assumes five valence protons and six valence neutrons
above a frozen $^{16}$O core. The total dimension of the many-body
space is 80,115, but this is constructed using only 792 five-proton
states and 923 six-neutron states.

The  reader will note that $792 \times 923 = 731016 \gg 80115$.
Indeed, not every five-proton state can be combined with every
six-neutron state. The restriction is due to conserving certain additive quantum
numbers, and this restriction turns out to limit usefully the nonzero
matrix elements of the many-body Hamiltonian, which we will discuss more in the
next section.

For our example, we chose  total $M = + 1/2$ (though we could have
chosen a different half-integer value). This basis requires that
$M_p + M_n = M$; and for some given $M_p$, \textit{every} proton
Slater determinant with that $M_p$ combines with \textit{every}
neutron Slater determinant with $M_n = M - M_p$. This is illustrated
in Table \ref{al27}, which shows how the many-body basis is
constructed from 792 proton Slater determinants and 923 neutron
Slater determinants. Note we are ``missing'' a neutron Slater
determinant; the lone $M_n = -7$ state has no matching (or
`conjugate') proton Slater determinants.

As a point of terminology, we divide up the basis (and thus any wavefunction vectors)
into \textit{sectors}, each of which is labeled by $M_p$, and any additional quantum numbers
such as parity $\Pi_p$; that is, all the basis states constructed with the same
$M_p$ ($\Pi_p$, etc.) belong to the same basis `sector' and have contiguous indices.
Basis sectors are also useful for grouping operations of the Hamiltonian, as described
below, and can be the basis for distributing vectors across many processors, although
because sectors are of different sizes this creates nontrivial issues for load balancing.

\begin{figure}
\includegraphics [width = 7.5cm,clip]{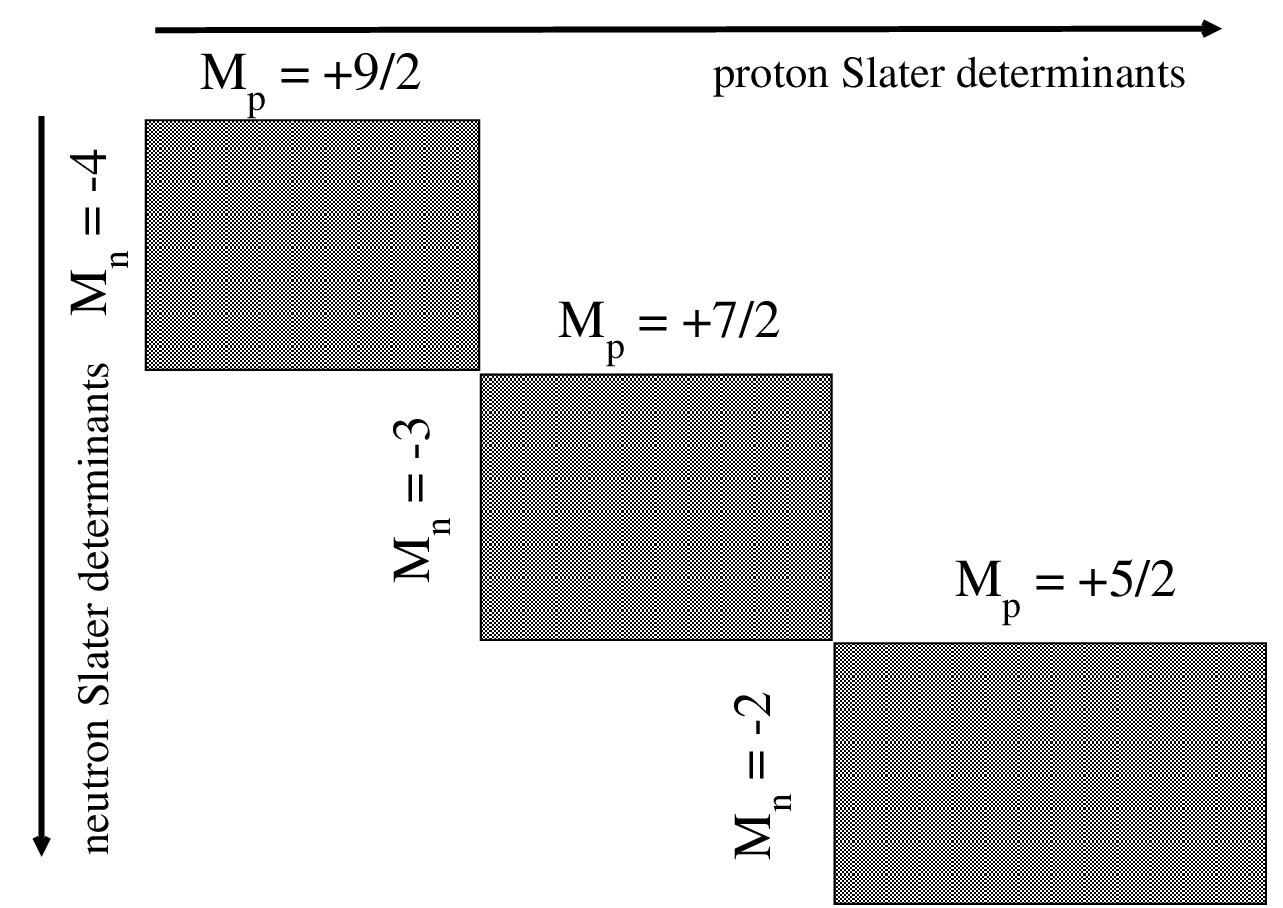}
\caption{\label{factor_basis} Illustration of factorization of the
$M$-scheme many-body basis. Along the $x$-axis we have proton Slater determinants ordered
by $M_p$, while along the $y$-axis we have neutron Slater determinants also
ordered by $M_n$ (but in reverse order). Any proton Slater determinant with
given $M_p$ will combine with any and all neutron Slater determinants with
conjugate $M_n = M -M_p$.  Each block therefore represents a \textit{sector} of the
basis in our terminology.
This example is for $M = 1/2$, taken from the text.
}
\end{figure}

This factorization is further illustrated in
Fig.~\ref{factor_basis}; we can think of the many-body basis being
formed by rectangles, the sides of which are sectors of the proton
and neutron Slater determinants, organized by $M_p$ and $M_n$.
Again, one can generalize this to other additive/multiplicative
quantum numbers such as parity: in multi-shell calculations the
total parity is fixed with $\Pi = \Pi_p \times \Pi_n$.

\begin{table}
\caption{Decomposition of the $M$-scheme basis for 5 protons and 6
neutrons in the $sd$ valence space ($^{27}$Al), with total $M = M_p
+ M_n +1/2$. Here ``pSD'' = proton Slater determinant and ``nSD'' =
neutron Slater determinant, while ``combined'' refers to the
combined proton+neutron many-body basis states.  The subset of the
basis labeled by fixed $M_p$ (and thus fixed $M_n$) we label a
'sector' of the basis. \label{al27}}
\begin{tabular}{|r|r|r|r|r|}
\hline
$M_p$ & \# pSDs  & $M_n$ & \# nSDs & \# combined \\
\hline
+13/2 & 3 & -6 & 9 & 27 \\
+11/2 & 11 & -5 & 21 & 231 \\
+9/2 & 28 & -4 & 47 & 1316\\
+7/2 & 51 & -3 & 76 & 3876 \\
+5/2 & 80 & -2 & 109 & 8720 \\
+3/2 & 104 & -1 & 128 & 13,312\\
+1/2 & 119 &  0 & 142 & 16,898\\
-1/2 & 119 &  +1 & 128 & 15,232 \\
-3/2 & 104 &  +2 & 109 & 11,336\\
-5/2 & 80 &  +3 & 76 & 6080\\
-7/2 & 51 &  +4 & 47 & 2444\\
-9/2 & 28 &  +5 & 21 & 588\\
-11/2 & 11 &  +6 & 9 & 99 \\
-13/2 & 3 &  +7 & 1 & 3 \\
\hline
Total & 792 & & 923 & 80,115 \\
\hline
\end{tabular}
\end{table}

The factorization leads to an impressive compactification of the
basis. We could explicitly write down each of the  80,115 basis
states in terms of their component proton and neutron Slater
determinants,  that is,
\begin{equation}
| \alpha \rangle = | \mu_p \rangle | \nu_n \rangle.
\end{equation}
Here we index the $M$-scheme basis  by $\alpha = 1, \ldots,80115$, while
the proton Slater determinants are indexed by
$\mu_p = 1,\ldots,792$  and
the neutron Slater determinants by $\nu_n = 1,\ldots, 923$.
It is straightforward to construct index functions $f_p, f_n$ 
such that \cite{ANTOINE,NATHAN}
\begin{equation}
\alpha = f_p(\mu_p) + f_n(\nu_n)
\label{basis_index}
\end{equation}

This is an example of factorization. Instead of storing explicitly
each and every basis state, one only needs the much smaller set of
proton and neutron Slater determinants, and the indexing functions
to map to the combined many-body basis.  Table \ref{basis_factor} gives
the number of component proton and neutron Slater determinants for a
number of representative cases.

One can also introduce some useful truncations of the many-body basis, also
based upon additive weights that act like quantum numbers.
In order not to muddy the waters, we give a description of a specific scheme
in  \ref{weighting}.

\begin{table}
\caption{Factorization of the $M$-scheme basis ($M = 0$ for even $A$, or $1/2$ for odd $A$) for
selected atomic nuclei in terms of the number of proton and
neutron Slater determinants needed.
The model spaces are described in Appendix B.
\label{basis_factor}}
\begin{tabular}{|c|c|c|c|c|c|}
\hline
Nuclide  &  space  & basis  & proton & neutron & $\#$ basis \\
         &         &  dim.  &  SDs        &  SDs & sectors \\
\hline
$^{28}$Si  & $sd$ & $9.4 \times 10^4$ & 924  &  924 & 15 \\
$^{56}$Fe  & $pf$ & $5.0 \times 10^8$ & 38760 & 184,722 & 27 \\
$^{56}$Ni  & $pf$ & $1.1 \times 10^9$ & $1.2 \times 10^5$ &  $1.2 \times 10^5$ & 29 \\
$^{6}$He   & $N_\mathrm{shell} =6$ & $1.4 \times 10^{9}$ & 6216 & $6 \times 10^6$ & 42 \\
$^{4}$He   & $N_\mathrm{shell} =10$ & $2.7 \times 10^{8}$ & 96,580 &  96,580 & 74 \\
$^{9}$Li   & $N_\mathrm{max} =10$ & $3.5\times 10^{8}$ & $1.5 \times 10^{5}$ & $1.4 \times 10^7$ & 126 \\
$^{9}$Be   & $N_\mathrm{max} =10$ & $5.7\times 10^{8}$ & $1.1 \times 10^{6}$ & $5.1 \times 10^6$ & 136\\
$^{4}$He   & $N_\mathrm{max} =22$ & $8.6 \times 10^{7}$ & $3 \times 10^{5}$ & $3 \times 10^5$ &  307 \\
$^{14}$C   & $N_\mathrm{shell} =3$ & $2.6 \times 10^{8}$ & $4\times 10^4$ &  $1 \times 10^5$ & 34  \\
$^{12}$C   & $N_\mathrm{shell} =4$ & $5.9 \times 10^{11}$ & $4\times 10^6$ &  $4\times 10^6$ & 58 \\
$^{12}$C   & $N_\mathrm{max} =8$ & $5.9 \times 10^{8}$ & $5 \times 10^6$ & $5 \times 10^6$ & 105 \\
\hline
\end{tabular}
\end{table}

Now that we have introduced the concept of factorization for the basis, we
turn to its usage for the interaction.

\section{The Hamiltonian and its factorization}

\label{factorizeHamiltonian}

In second quantization, a Hamiltonian takes the form
\begin{equation}
\hat{H} = \sum_{ij} T_{ij} \hat{a}^\dagger_i \hat{a}_j
+ \frac{1}{4} \sum_{ijkl} V_{ijkl} \hat{a}^\dagger_i \hat{a}^\dagger_j
\hat{a}_l \hat{a}_k
\end{equation}
where $T_{ij}$ are the one-body Hamiltonian matrix elements, which may include kinetic energy
and some external or mean-field potential, and $V_{ijkl}$, the two-body
matrix elements; as the latter is the primary computational challenge, we
ignore the one-body part.  One can also add three-body terms, etc., but the
underlying principals are unchanged.

The matrix elements are not uncorrelated, due to hermiticity, fermion antisymmetry,
and, most germane to this discussion, rotational invariance.
The details can be found in standard monographs, e.g. \cite{BG77}.  For our purposes,
we only care about selection rules which arise from additive or multiplicative
quantum numbers. What this means is, for example, the
requirement that, unless
$$
m_i + m_j - m_k -m_l =0
$$
then $V_{ijkl} = 0$. This is an example of a conservation law/selection rule.  A similar selection rule exists for parity. Below we discuss how we can
exploit this for two-species systems, but first, we discuss \textit{why}
the $M$-scheme Hamiltonian matrix is so sparse, and, furthermore, why the
nonzero matrix elements have enormous redundancy.

\subsection{Sparsity and redundancy of the matrix elements}
\label{sparsity}

Table I illustrated the sparsity of $M$-scheme Hamiltonian matrices.
This can be understood as arising because we restrict ourselves to a
two-body Hamiltonian.  We expand upon this idea here.

As explained above, one can represent the basis Slater determinants
as binary numbers, with a `1' representing an occupied state and a
`0' an occupied state.  The action of a destruction operator
$\hat{a}_i$ is to replace a `1' in the $i$th position with a `0',
while a creation operator $\hat{a}^\dagger_i$ does the opposite:
\begin{eqnarray}
\hat{a}_4 | 01011100 \rangle = -| 01001100 \rangle; \\
\hat{a}^\dagger_4 | 01001100 \rangle = -| 01011100 \rangle; \\
\hat{a}_4 | 01001100 \rangle = 0; \hat{a}^\dagger_4 | 01011100
\rangle = 0.
\end{eqnarray}
(The phases arise because of fermion anticommutation relations: every time an
annihilation operator anticommutes past a creation operator, we pick up a minus sign.)
Trying to create a particle where there already is one gives a zero amplitude; this is
the Pauli exclusion principle for fermions. Conversely, trying to destroy a particle where
none exists also gives a zero amplitude.

Thus, the action of the operator $\hat{a}^\dagger_i \hat{a}^\dagger_j \hat{a}_l
\hat{a}_k $ is to destroy particles in states $k$ and $l$ and put particles in
states $i$ and $j$. To simplify, let us assume $i,j,k$, and $l$ are all distinct.
Then the amplitude of this operator between two Slater determinants
$\mu$ and $\mu^\prime$,
$\langle \mu^\prime | \hat{a}^\dagger_i \hat{a}^\dagger_j \hat{a}_l
\hat{a}_k | \mu \rangle$ is zero unless:

\noindent $\bullet$ The states $k$, $l$ are occupied in $| \mu \rangle$ and
unoccupied in $| \mu^\prime \rangle$ ;

\noindent $\bullet$ The states $i$, $j$ are occupied in $| \mu^\prime \rangle$
and unoccupied in $| \mu\rangle$; and

\noindent $\bullet$ all other particles in  $| \mu \rangle$ and
 $| \mu^\prime \rangle$ occupy identical states; these are  \textit{spectators}
and do not change as one goes from $\mu$ to $\mu^\prime$.

As one might imagine, these stringent conditions are difficult to meet; hence,
the sparsity found in Table I.

Now the story is not done. Not only is the Hamiltonian matrix
sparse, the nonzero matrix elements are furthermore highly redundant. In the action of
the Hamiltonian, the operator $\hat{a}^\dagger_i \hat{a}^\dagger_j
\hat{a}_l \hat{a}_k $ has a numerical amplitude $V_{ijkl}$. Now this
operator will be able to connect many dozens of different pairs of
Slater determinants, leading to many dozens of different many-body
Hamiltonian matrix elements. But only the spectators
differ, while the value of the matrix element $V_{ijkl}$, up to some
phase, remains the same.

This leads to large redundancies, as illustrated in Table \ref{redundancy},
from factors spanning from 20 to $10^6$.
In a two-species system, one can devise a factorization algorithm to take
 partial advantage of this redundancy to reduce the
memory load, as described in the following subsections.

\begin{table}
\caption{Some model spaces for atomic nuclei and their  $M$-scheme ($M=0$) dimensions and
the average redundancy of the Hamiltonian matrix elements (m.e.s), defined
as the ratio of the number of nonzero matrix elements (excluding Hermitian
conjugates) to the number of
\textit{unique} matrix elements.
The model spaces are described in Appendix B.
\label{redundancy}}
\begin{tabular}{|c|c|c|c|c|c|}
\hline
Nuclide  &  space  & basis  & $\#$ nonzero & $\#$ unique & average \\
         &         & dim    &  m.e.s  & m.e.s & redundancy\\
\hline
$^{28}$Si  & $sd$ & $9.4 \times 10^4$ & $2.6 \times 10^7$  & 2800  & 9300 \\
$^{52}$Fe  & $pf$ & $1.1 \times 10^8$ & $8.9 \times 10^{10}$ & $2.2\times 10^4$ & $4 \times 10^6$ \\
$^{56}$Ni  & $pf$ & $1.1 \times 10^9$ & $1.2 \times 10^{12}$ & $2.2\times 10^4$ &
$5 \times 10^7$ \\
$^{4}$He   & $N_\mathrm{shell} =8$ & $2.9 \times 10^7$ & $1.8 \times 10^{11}$ &
$1.4 \times 10^8$ & 1300 \\
$^{4}$He   & $N_\mathrm{max} =16$ & $6 \times 10^6$ & $2.5 \times 10^{10}$ &  $1.5 \times 10^9$  & 18  \\
$^{12}$C   & $N_\mathrm{shell} =3$ & $8.2 \times 10^7$ & $5.2 \times 10^{10}$   & $1.6 \times
10^4$   & $3 \times 10^6$  \\
$^{12}$C   & $N_\mathrm{max} =8$ & $6 \times 10^8$ & $6.4\times 10^{11}$ & $5 \times 10^7$ & $1\times 10^4$ \\
\hline
\end{tabular}
\end{table}

\subsection{The Hamiltonian with two species}

\label{Htwospecies}

With two species, the Hamiltonian can be
written  as a sum of different parts
\begin{equation}
\hat{H} = \hat{H}_x + \hat{H}_y + \hat{H}_{xx} +\hat{H}_{yy} + \hat{H}_{xy}.
\end{equation}
Here $\hat{H}_x$ is a one-body operator, \textit{i.e.}, kinetic
energy plus external potential, that acts only on species
\textit{x}, $\hat{H}_{xx}$ is a two-body operator which denotes the
interaction between two particles of species \textit{x}, and
$\hat{H}_{xy}$ is a two-body operator that denotes the interaction
between one particle of species \textit{x} and one particle of
species \textit{y}; the generalizations to species $y$, $\hat{H}_y$ and
$\hat{H}_{yy}$, are analogous.  The one-body terms are easy to work
with, so, henceforth, we focus exclusively on the two-body
interactions.

We write
\begin{equation}
\hat{H}_{xx} = \frac{1}{4} \sum_{ijkl}V^{(xx)}_{ijkl}\hat{a}^\dagger_i(x) \hat{a}^\dagger_j(x)
\hat{a}_l(x)\hat{a}_k(x). \label{Hxx}
\end{equation}
The analogous operator for species $y$,  $\hat{H}_{yy} $ has a similar form. The cross-species interaction, that
is interaction between a particle of species $x$ and a particle of species $y$, is
\begin{equation}
\hat{H}_{xy} =  \sum_{ijkl}V^{(xy)}_{ijkl}\hat{a}^\dagger_i(x) \hat{a}^\dagger_j(y)
\hat{a}_l(y)\hat{a}_k(x), \label{Hxy}
\end{equation}
which more broadly can be expanded in a factorized fashion:
\begin{equation}
\hat{H}_{xy} = \sum_{ijkl}V^{(xy)}_{ijkl}  \hat{\cal O}^{(x)}_{ik} \hat{\cal O}^{(y)}_{jl}
\label{h_xy}
\end{equation}
where, for example,
\begin{equation}
\hat{\cal O}^{(x)}_{ij} =   \hat{a}^\dagger_i(x) \hat{a}_j(x).
\label{1bodyop}
\end{equation}
Because the model space is finite all such expansions are also finite.

Now, we can sketch out the Hamiltonian matrix element (again, considering only
the two-body interactions)
\begin{eqnarray}
\langle  \nu^\prime_y | \langle \mu^\prime_x |
\hat{H}| \mu_x \rangle | \nu_y \rangle
=
 \langle \mu^\prime_x|
\hat{H}_{xx}| \mu_x \rangle  \delta_{\nu^\prime \nu} +
\langle \nu^\prime_y|
\hat{H}_{yy}| \nu_y \rangle  \delta_{\mu^\prime \mu} + \\
\sum_{ab} V_{ab}
 \langle \mu^\prime_x|\hat{\cal O}^{(x)}_a | \mu_x \rangle
 \langle \nu^\prime_y|\hat{\cal O}^{(y)}_b | \nu_y \rangle \nonumber
\end{eqnarray}
This allows us to begin to see the route to factorization and its efficiency.
To begin with, the component matrix elements,
$ \langle \mu^\prime_x|
\hat{H}_{xx}| \mu_x \rangle ,
\langle \mu^\prime_x|\hat{\cal O}^{(x)}_a | \mu_x \rangle $,
etc.,  are much smaller in number than the full matrix elements, as
conservation laws/selection rules dramatically restrict the number and
coupling of these components.

We assume the Hamiltonian to be rotationally
invariant, which means that the Hamiltonian is an angular momentum
scalar. (Notice that we also assume the number of particles of each species
is conserved,
 a condition that  could be relaxed although it would lead to additional complications.)
 Because
the Hamiltonian commutes not only with $\hat{J}^2$ but also $\hat{J}_z$, this means
that the eigenvalue of $\hat{J}_z$, or $M$, is conserved.
Earlier, we discussed how this allowed us to invoke a fixed-$M$ basis, which is
easy to construct using Slater determinants.  But now we go further: because
the interaction cannot change $M$,
in Eqs.~(\ref{Hxx}) and (\ref{Hxy}) we have
\begin{equation}
m_i + m_j - m_k - m_l = 0.
\end{equation}
Similarly for parity,
\begin{equation}
\pi_i \times \pi_j \times \pi_k \times \pi_l = +1.
\end{equation}
The conservation of these quantum numbers dramatically restricts the
number and the coupling of the matrix elements, as we will now lay out.


\subsection{Two-body jumps}

\begin{figure}
\includegraphics [width = 7.5cm,clip]{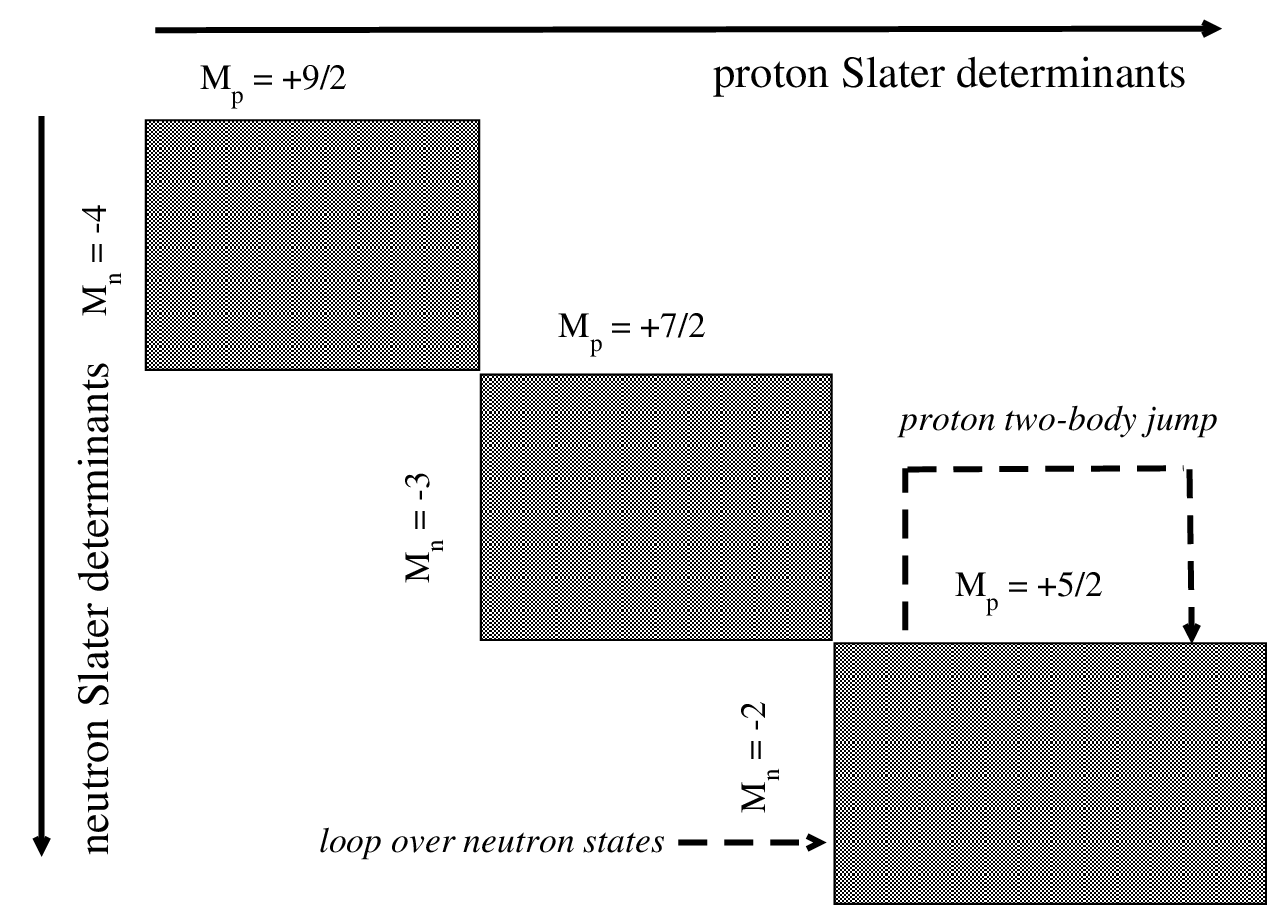}
\caption{\label{factor_Hpp} Illustration of factorization of $H_{pp}$ for
the example in the text. Because the conjugate neutron Slater determinant
does not change, the `two-body jump' cannot change either $M_n$ or $M_p$.
Instead we loop over all (proton) two-body jumps and also loop over all conjugate
(neutron) Slater determinants.
}
\end{figure}

Consider $\hat{H}_{xx}$, the interaction between two particles
of species $x$; $\hat{H}_{yy}$ works exactly the same.
As described above,
\begin{equation}
\label{Hxx_factor}
\langle \nu^\prime_y | \langle \mu^\prime_x |
\hat{H}_{xx}| \mu_x \rangle | \nu_y \rangle
=
 \langle \mu^\prime_x |
\hat{H}_{xx}| \mu_x \rangle  \delta_{\nu^\prime \nu} .
\end{equation}
What does this mean for computing the matrix element of $\hat{H}_{xx}$?  Because
$\nu = \nu^\prime$ for the $y$-basis component, $\hat{H}_{xx}$ cannot
change $M_y$ or $\Pi_y$, which in turn means that it cannot change $M_x$ or
$\Pi_x$.  Figure \ref{factor_Hpp} gives a
schematic version.

Thus, although we need the
$ \langle \mu^\prime_x |\hat{H}_{xx}| \mu_x \rangle $,
we only need them diagonal in quantum numbers such as $M_x$, which enormously
reduces the amount of information. The $ \langle \mu^\prime_x |\hat{H}_{xx}| \mu_x \rangle $ are called `two-body jumps.'
Each two-body jump consists of the following information:

\noindent $\bullet$ the initial $x$-state label $\mu$;

\noindent $\bullet$ the final $x$-state label $\mu^\prime$;

\noindent $\bullet$ and the value of the matrix element
$ \langle \mu^\prime_x|\hat{H}_{xx}| \mu_x \rangle $ (which includes a phase $\pm 1$ from
fermion anticommutation).

These `jumps' can be stored as simple arrays, along with indexing information
that tells us the start and stop points for two-body jumps for a given
$M_x$ etc.  Because the jumps are at the heart of the factorization
algorithm, they are best kept as simple arrays; storing within derived types
in modern Fortran, for example, can increase run time by a factor of two.

The application of $\hat{H}_{xx}$ is given below in pseudocode. 
It is helpful to remind ourselves of what we
learned about the factorization of the basis in Section \ref{factor_mscheme}:
any $x$-basis state $| \mu_x \rangle $ has quantum numbers
$M_x$, $\Pi_x$  (and optionally the `weighting' $W_x$, used for many-body truncations, as discussed 
in \ref{weighting}); the subset of $x$-states $| \mu_x \rangle$ with the same quantum 
numbers is called a sector, which in the pseudocode we label by $\Gamma_x$. 
A $y$-sector $\Gamma_y$ is `conjugate' to 
$\Gamma_x$ if $M_x + M_y = M$, the fixed total $J_z$ for the calculation, and $\Pi_x \times 
\Pi_y = \Pi$ and $W_x + W_y \leq W_\mathrm{max}$.  
 Furthermore, as in Eqn.~\ref{basis_index} there exist 
index function $f_x(\mu_x)$ and $f_y(\nu_y)$ whose sum gives us the 
index $\alpha$ of the combined basis state.

Then the pseudocode is:

\medskip

\parbox{11.0cm}
{ FOR all $x$-sectors $\Gamma_x$ 

\smallskip

\hfill \parbox{10.5cm}
     { FOR  all the $x$-species two-body jumps $\in \Gamma_x$

\smallskip

\hfill \parbox{10.0cm}
           { Fetch  $\mu_x$, $\mu_x^\prime \in \Gamma_x$ and

            Fetch matrix element  
            $\langle \mu^\prime_x| \hat{H}_{xx}| \mu_x\rangle$  from the jumps arrays

\smallskip

            FOR all $\Gamma_y$  conjugate to $\Gamma_x$

\hfill \parbox{9.5cm}
                { FOR all $\nu_y \in \Gamma_y$

\smallskip

\hfill \parbox{9.0cm}
                 {  $\alpha_i = f_x(\mu_x) + f_y(\nu_y)$

                  $\beta_f = f_x(\mu^\prime_x) + f_y(\nu_y)$ 

                  $v_f(\beta_f) = v_f(\beta_f) + v_i(\alpha_i) \times \langle \mu^\prime_x|\hat{H}_{xx}| \mu_x\rangle $ 

$v_f(\alpha_i) =  v_f(\alpha_i) + v_i(\beta_f) \times \langle 
\mu^\prime_x|\hat{H}_{xx}| \mu_x \rangle^*$ (hermiticity) 
                   }

\smallskip
             END FOR}
\smallskip

            END FOR}

\smallskip

      END FOR}

\smallskip

END FOR
}

\medskip

In a purely serial code lookup of the jump
information is time consuming, and so looping over the spectator
$y$-states is best as the innermost loop, but for example when using
OpenMP or other shared-memory parallel schemes, we use the loop over
the spectator $y$-states the outer loop in order to avoid data
collisions.

To illustrate this algorithm, we return to the example given in
Section \ref{factor_mscheme},  and in particular in Table
\ref{al27}. The proton-proton or $xx$ part of the Hamiltonian cannot
change $M_n$ and so also cannot change $M_p$. Now let's consider the
specific case of the sector of the basis with $M_p = +5/2$ and $M_n
= -2$. There are a total of 8720 basis states in this sector, so
there are possibly $8720^2 =68 \times 10^6$ matrix elements.
However, $\hat{H}_{pp}$ cannot change the neutron Slater
determinant, so that, as in Eq.~(\ref{Hxx_factor}), $\nu =
\nu^\prime$.

Now in this sector, there are 80 proton Slater determinants $|\mu_p \rangle$
with $M_p = 5/2$, so that in principle there could be $80^2 = 6400$ matrix elements
$ \langle \mu^\prime_p|\hat{H}_{pp}| \mu_p \rangle $  (or half that
if we take into account hermiticity). But we have a two-body interaction and five
particles, so that for each nonzero matrix element there are always three static spectators.
This reduces the number of nonzero matrix elements of
$ \langle \mu^\prime_p|\hat{H}_{pp}| \mu_p \rangle $
to 2677.

With the two-body jumps in hand, we can easily reconstruct the full matrix
element, Eq.~(\ref{Hxx_factor}), for this sector of the basis.  We loop over all 2677 two-proton
jumps, which takes us from one proton Slater determinant  $| \mu_p \rangle$
to another $|\mu^\prime_p \rangle$, including the value of
the matrix element. We also have a fast, inner loop
over the 109 neutron Slater determinants $| \nu_y \rangle$ which do not change.
(Note: if we parallelize our code with memory sharing, e.g. with OpenMP, it is
useful to either make the loop over $\nu_y$ the outer loop 
or to order the jumps based on the index of the final state 
and sort on thread boundaries 
in order to avoid data collisions.)
We invoke the straightforward indexing of the basis, Eq.~(\ref{basis_index}), to
obtain the indices of the initial and final basis states. End result: out of
a possible 68 million matrix elements in this sector, we use just 2677 two-proton jumps to
find the $2677 \times 109 = 291793$ nonzero matrix elements (a sparsity of $0.4\%$).
(Furthermore, there are only 350 unique values of the matrix elements.)  This illustrates
how factorization can compress, without approximation, an already very sparse matrix
into a relatively small amount of memory.

\begin{table}
\caption{Number of one- and two-body `jumps' and storage requirements for representative
atomic nuclei in different model spaces (described in Appendix B). For storage
of nonzero matrix elements (penultimate column) we assume each many-body matrix element is stored
by a 4-byte real number and its location encoded by a single 4-byte integer.  Storage of a single
jump (initial and final Slater determinants for a species, and matrix element and phase) requires
13 bytes.  All storage (final two columns) are in gigabytes (GB). 
\label{jumps}}
\begin{tabular}{|c|c|c|c|c|c|c|}
\hline
Nuclide  &  space  & basis  &  $\#$ 1-body & $\#$ 2-body  & Store   & Store \\
         &         & dim    &  jumps & jumps  & m.e.s   & jumps \\
\hline
$^{28}$Si  & $sd$ & $9.4 \times 10^4$ & $4.8 \times 10^4$
& $7.6 \times 10^3$  & 0.2  & 0.002 \\
$^{52}$Fe  & $pf$ & $1.1 \times 10^8$ &  $4.0\times 10^6$ & $8.5 \times 10^6$ & 700 & 0.16 \\
$^{56}$Ni  & $pf$ & $1.1 \times 10^9$ &  $1.5\times 10^7$ &
$4.0 \times 10^7$ & 9800 & 0.6 \\
$^{4}$He   & $N_\mathrm{shell} =8$ & $2.9 \times 10^7$ &
$1.4 \times 10^7$ & $2.9 \times 10^7$ &  1500 &  0.6\\
$^{4}$He   & $N_\mathrm{max} =22$ & $9 \times 10^7$ &   $5.3 \times 10^8$  & $4.7 \times 10^9$ & 9300 &  69  \\
$^{12}$C   & $N_\mathrm{shell} =3$ & $8.2 \times 10^7$ &  $4 \times
10^6$   & $7 \times 10^6$  & 400 & 0.14 \\
$^{12}$C   & $N_\mathrm{shell} =4$ & $6 \times 10^{11}$ & $8\times 10^8$    & $2.4 \times 10^9$ & $10^7$ & 43 \\
$^{12}$C   & $N_\mathrm{max} =8$ & $6 \times 10^8$ &  $6 \times 10^8$ & $3 \times 10^9$
& 5200 & 45 \\
$^{13}$C   & $N_\mathrm{max} =6$ & $3.8 \times 10^{7}$ & $7 \times 10^{7}$ & $3 \times 10^8$ & 210 & 4.3 \\
\hline
\end{tabular}
\end{table}

Comparison between storage scheme and factorization is given in Table \ref{jumps} and\
discussed below in Subsection \ref{memory_required}.

\subsection{One-body jumps}

\begin{figure}
\includegraphics [width = 7.5cm,clip]{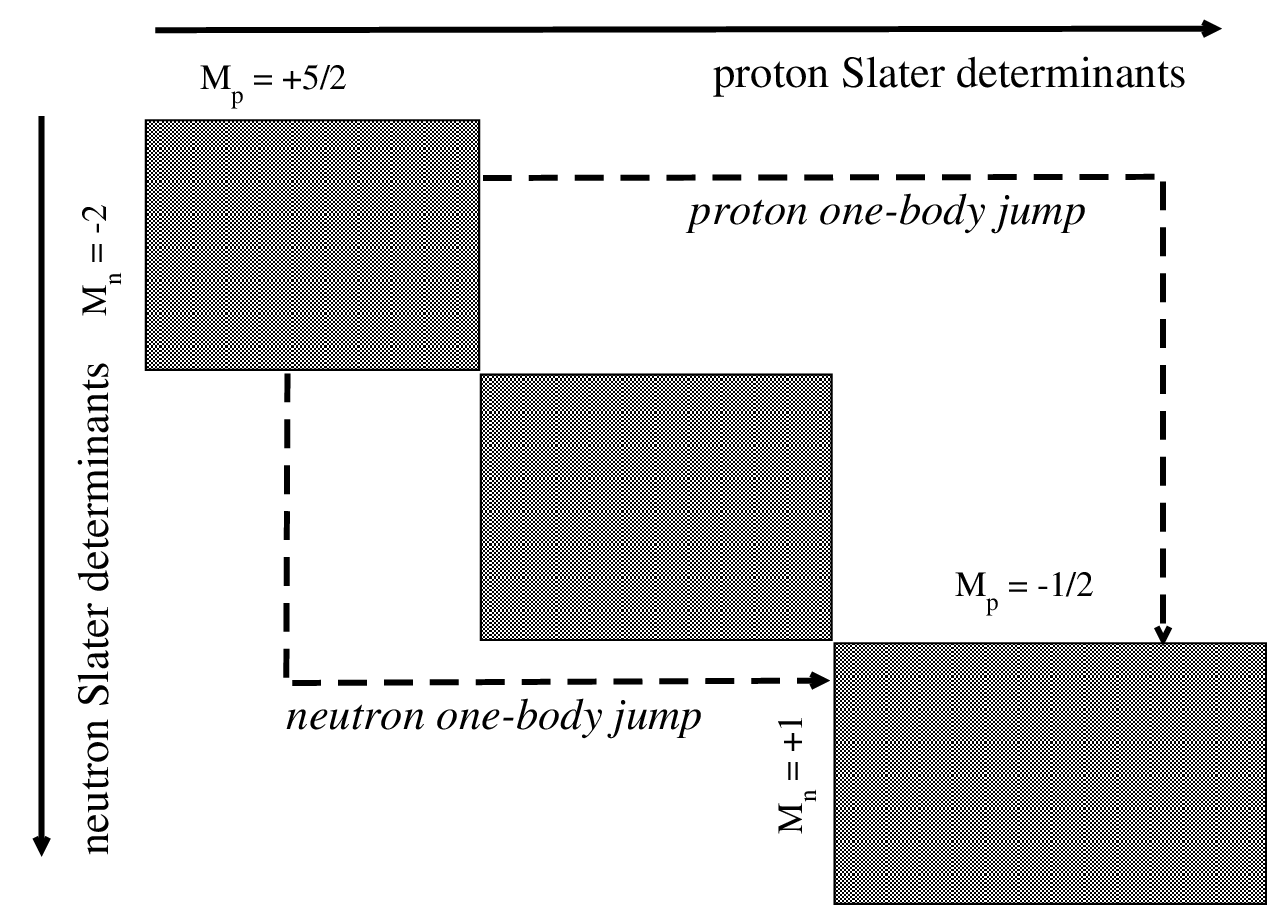}
\caption{\label{factor_Hpn} Illustration of factorization of $H_{pn}$ using
one-body jumps, for the example in the text. The change in $M_p$ must be matched
by the change in $M_n$.
}
\end{figure}

The action of $\hat{H}_{xy}$ is more complicated: it moves one particle of species
$x$ and one particle of species $y$. Nonetheless, factorization is still viable, and, in fact,
it greatly reduces the memory storage requirements.

Using Eq.~(\ref{h_xy}), the matrix element of $\hat{H}_{xy}$ is
\begin{equation}
\langle \nu^\prime_y | \langle \mu^\prime_x|
\hat{H}_{xy} | \mu_x \rangle | \nu_y \rangle
=
 \sum_{ab} V_{ab}
\langle \mu^\prime_x|
\hat{\cal O}^{(x)}_a  |\mu_x \rangle
\langle  \nu^\prime_y | \hat{\cal O}^{(y)}_b | \nu_y \rangle
\end{equation}
We call $\langle \mu^\prime_x| \hat{\cal O}^{(x)}_a | \mu_x \rangle
$ and $\langle \nu^\prime_y | \hat{\cal O}^{(y)}_b | \nu_y \rangle $
one-body jumps for species $x$ and $y$, respectively. Like two-body
jumps, for each one-body jump we must store the labels of the
initial and final Slater determinants plus the label $a$ of the
one-body operator $\hat{\cal O}_a$.

Because neither the $x$ nor the $y$ Slater determinants are spectators,  individual
quantum numbers such as $M_x$ and $M_y$ are no longer constant. On the other hand,
because the total $M = M_x + M_y$ \textit{is} constant, we know that $M_y$ must
change in an equal and opposite manner to the change in $M_x$.

So, going back to our example from the previous subsection (Table
\ref{al27}), if we start in the sector  $M_p = +5/2$, $M_n =
-2$, if the proton one-body jump takes us to $M_p = -1/2$ then we
\textit{must} have a conjugate neutron one-body jump that takes us
to $M_n = +1$.  Figure \ref{factor_Hpn} gives a schematic version.

For $H_{xx}$, we looped over the 2-body \textit{jumps} of species $x$ and then had a
simple loop over the allowed conjugate \textit{states} of species $y$ dictated by
quantum numbers. For $H_{xy}$, we must first identify the initial and
final $M$-scheme sectors, and loop over all the associated one-body $x$-jumps
\textit{and} all the conjugate one-body $y$-jumps. The pseudocode is:

\medskip

\parbox{11cm}
{
FOR all initial $\Gamma^i_x$, all final $\Gamma^f_x$

\smallskip

\hfill \parbox{10.5cm}
     {
      FOR all $\Gamma^i_y$ conjugate to $\Gamma^i_x$, $\Gamma^f_y$ conjugate to $\Gamma^f_x$

\smallskip

\hfill \parbox{10cm}
       {FOR all $x$-jumps $\in \Gamma^i_x \rightarrow \Gamma^f_x$, 
all $y$-jumps  $\in \Gamma^i_y \rightarrow \Gamma^f_y$,

\smallskip

\hfill \parbox{9.5cm}
               {Fetch $\mu_x \in \Gamma^i_x, \mu_x^\prime \in \Gamma^f_x$

\smallskip

               Fetch $\nu_y \in \Gamma^i_y, \nu_y^\prime \in \Gamma^f_y$

\smallskip

               Fetch $\langle \mu_x^\prime, \nu_y^\prime | H | \mu_x , \nu_y \rangle $

\smallskip

              $\alpha_i = f_x(\mu) + f_y(\nu)$;

\smallskip

              $\beta_f = f_x(\mu^\prime) + f_y(\nu^\prime)$

\smallskip

              $v_f(\beta_f) =v_f(\beta_f) + v_i(\alpha_i) \langle \mu_x^\prime, \nu_y^\prime | H | \mu_x \nu_y \rangle $

\smallskip

              $v_f(\alpha_i) = v_f(\alpha_i) + v_i(\beta_f) \langle \mu_x^\prime, \nu_y^\prime | H | \mu_x \nu_y \rangle^* $ (hermiticity)

              } 
\smallskip

      END FOR

              } 
\smallskip

      END FOR
        
      }
\smallskip

END FOR
}

\medskip

(Again: an $x$-sector, labeled $\Gamma_x$, is a subset of $x$-species Slater determinants with fixed quantum numbers $M_x$, $\Pi_x$, while  $\mu_x, \nu_y$ label $x$-species 
and $y$-species Slater determinants, respective, and $\alpha_i$ is the index of the many-body 
basis state constructed by the product of those two Slater determinants.)

Going from the subspace (or sector) $M_p = 5/2, M_n = -2$ (dimension 8720) to the
sector $M_p =-1/2$, $M_n = 1$ (dimension 15,232) the number of possible matrix elements is
$8720 \times 15,232 = 1.3\times 10^8$; but there are 337 proton
jumps and 419 neutron jumps and a total of $337 \times 419 = 132,823$ nonzero
matrix elements, with a sparsity of $0.1\%$.

Again, comparison between storage scheme and factorization is given in Table \ref{jumps} and\
discussed below in Subsection \ref{memory_required}.

\subsection{Three-body forces}
\label{3body}

One can generalize in a straightforward manner the previous algorithm to three-body forces.
Here, one now has $H_{xxx}$, $H_{xxy}$, $H_{xyy}$, and $H_{yyy}$.  The first and the last require
`three-body jumps,' while $H_{xxy}$ and $H_{xyy}$ require combining one- and two-body jumps.
In Table \ref{threebody}, we compare, for three-body forces, the sparsity as well as the
memory requirements to store the nonzero matrix elements and to store the jumps; these can
be compared to data in Tables \ref{nuclear_sparsity} and \ref{jumps}.  The matrices are, as expected,
significantly less sparse, while we retain the significant memory efficiency by storing jumps rather
than storing the nonzero matrix elements.

\begin{table}
\caption{Storage requirements for three-body interactions for representative
atomic nuclei in different model spaces (described in Appendix B). The corresponding
data for two-body interactions can be found in Tables \ref{nuclear_sparsity} and \ref{jumps}.
All storage (final two columns) are in gigabytes (GB). 
\label{threebody}}
\begin{tabular}{|c|c|c|c|c|c|}
\hline
Nuclide  &  space  & basis  &   Sparsity &  Store   & Store \\
         &         & dim    &             & m.e.s    & jumps \\
\hline
$^{28}$Si & $sd$ & $9 \times 10^4$ & 0.06 &  2.1 &  0.01 \\
$^{52}$Fe & $pf$ & $1 \times 10^8$  &  $3 \times 10^{-4}$ & $1.5 \times 10^4$ & 1.6 \\
$^{4}$He & $N_\mathrm{shell} = 8 $ & $3 \times 10^7$  & 0.03 &  $1.2 \times 10^5$ & 11 \\
$^{4}$He & $N_\mathrm{max} = 14 $ & $2 \times 10^6$  & 0.10 &  2000 & 16 \\
$^{12}$C & $N_\mathrm{shell} = 3 $ & $8 \times 10^7$  & $3\times 10^{-4}$ &  8800 & 1.4 \\
$^{13}$C & $N_\mathrm{max} = 6 $ &  $4 \times 10^7$  & $1 \times 10^{-3}$ & 6200   & 80 \\
$^{12}$C & $N_\mathrm{max} = 8 $ &  $6 \times 10^8$  & $1.5 \times 10^{-4}$ & $2 \times 10^5$   & 1100 \\

\hline
\end{tabular}
\end{table}

\subsection{Comparison of memory requirements}
\label{memory_required}

Using `jumps' to store the information about the many-body
Hamiltonian matrix elements is significantly more efficient than
storing the matrix elements, as illustrated for two-body
interactions in Table \ref{jumps}: the memory requirements are 50 to
100 times less for factorization, an enormous savings. Furthermore,
reconstruction of the matrix elements is very fast; in timing tests
it is no more than a factor of two slower than simply fetching the
matrix element from memory, and that is for $H_{xy}$; for
$H_{xx/yy}$ the timing between factorization and storage is similar.
Table \ref{threebody} makes the comparison for three-body forces,
where the ratio is even larger.

There are other advantages.  Setup of the arrays is very fast,
especially if one takes the method further as we describe in the
next section. Because we can `forecast' the number of operations
without actually creating or directly counting the nonzero matrix
elements, we can quickly obtain that information and distribute the
work for parallel processing, as we describe in Section
\ref{parallel}.

The take-away lesson is: while the $M$-scheme Hamiltonian matrix is
very sparse, one can dramatically reduce the storage requirements
further by one or two orders of magnitude using factorization; in the case of
three-body forces, the reduction can be as high as three orders of magnitude.  With
factorization one can run efficiently on a desktop machine problems
that would otherwise require storing the many-body Hamiltonian on
disk, with slow I/O, or distributing the matrix elements across many
compute nodes on a distributed memory parallel machine. Alternately,
on the \textit{same} parallel machine one could tackle a much larger
problem--50 or 100 times larger, even more in the case of three-body
forces as discussed  in Subsection \ref{3body} -- than with a
matrix storage scheme. Thus, despite the higher intrinsic complexity
of the algorithm (and we note that leadership class codes using
matrix storage are by no means `simple'), factorization can, using the same computational
resources, 
push the limits of calculations further.

\section{Factorization: the next step}
\label{haiku}

Inspired by ANTOINE \cite{ANTOINE}, our first attempt at on-the-fly
CI code was REDSTICK (unpublished), initiated while two of us (CWJ
and WEO) were at Louisiana State University in Baton Rouge, La. When
we began to plan our next generation code, BIGSTICK, we first looked
at the computational bottlenecks in REDSTICK. They were, in
decreasing order of importance:

\noindent $\bullet$ Inefficient application of the jumps when making
truncations on the many-body space (briefly: use of the $W$
weighting index to truncate the many-body space, as described in
 \ref{weighting}, further restricts application of the
jumps, but this was not handled in an efficient manner);

\noindent $\bullet$ Inefficient distribution of the matvec
(matrix-vector multiply) operations across parallel computational
nodes;

\noindent $\bullet$ Inefficient generation of the basis, especially
for large relative truncation based on weighting $W$; and

\noindent $\bullet$ Inefficient generation of the one- and two-
(and, for three-body interactions, three-) body jumps. Note that no
other on-the-fly CI code has 3-body capabilities; this was our major
motivation for writing REDSTICK and BIGSTICK.

The first bottleneck, inefficient application of the jumps, was addressed by organizing both the basis and the jumps by sectors, that is, labeling
by $M, \Pi$ and $W$.
The second bottleneck, inefficient parallelization, we address below in Section \ref{parallel}.  The final two bottlenecks
we addressed by introducing a new level of factorization.

As described above, the basis is factorized into a product of $x$- and
$y$-species Slater determinants (e.g. proton and neutron Slater determinants, or up- and down-spin electron 
Slater determinants) which have conjugate quantum numbers.
We have taken this strategy to the next level. Each Slater
determinant for given species is itself written as a product of two `half-Slaters,' one
constructed from single-particle states with $m < 0$, and the other
from single-particle states with $m \geq 0$.
\begin{equation}
| \mu_x \rangle  = | \omega^{(L)}_x  \rangle
| \omega^{(R)}_x \rangle .
\end{equation}
Here `L' denotes a `left' half-Slater determinant (hSD) , constructed with single-particle
states with $m < 0$,  `R' denotes a `right' hSD constructed with
single-particle states with $m \geq 0$\footnote{We use $m=0$ for some atomic physics calculations, when 
the spin quantum number $m_{\uparrow,\downarrow}$ is associated with the two species, and thus otherwise $m$ is associated 
with orbital angular momentum $L$ with integer values. }, and $\omega^{(L), (R)}$ 
denotes the  quantum
numbers associated with each hSD, notably $M^{(L), (R)}$, $\Pi^{(L),(R)}$,  $W^{(L),(R)}$, and, crucially, the number of particles $n^{(L),(R)}$.
As usual,
$M_x = M^{(L)} + M^{(R)}$, etc, while $n_x = n^{(L)} + n^{(R)}$.  Any full basis state is now
the product of four half-Slater determinants, e.g. $| \alpha \rangle =
| \omega^{(L)}_x  \rangle
| \omega^{(R)}_x \rangle | \omega^{(L)}_y  \rangle
| \omega^{(R)}_y \rangle$.   However this combining is only done implicitly and never explicitly.

Crucially, and unlike the Slater determinants of species $x$ and $y$,
these half-Slater determinants
do not have fixed particle number, which now acts as a new quantum number.
While this adds a layer of complexity, it also offers several advantages. In the same
way one constructs a very large total basis from much smaller lists of $x$- and $y$-
Slater determinants, the number of required hSDs is much smaller still.
For any given set of quantum numbers $n$, $M$, $\Pi$ (and $W$), the list of half-Slaters
is  small.  Examples are given in Table \ref{haiku_factor}; one gets bigger
savings for more particles and for full-configuration bases.

\begin{table}
\caption{Factorization of the $M$-scheme basis ($M = 0$) for
selected atomic nuclei in terms of the number of 'half-Slater
determinants'.
\label{haiku_factor}}
\begin{tabular}{|c|c|c|c|c|}
\hline
Nuclide  &  space  & basis  & proton & proton  \\
         &         &  dim.  &  SDs        &  hSDs \\
\hline
$^{28}$Si  & $sd$ & $9.4 \times 10^4$ & 924  &  128 \\
$^{52}$Fe  & $pf$ & $1.1 \times 10^8$ & 38760 & 1696 \\
$^{56}$Ni  & $pf$ & $1.1 \times 10^9$ & $1.2 \times 10^5$ & 2026 \\
$^{4}$He   & $N_\mathrm{shell} =8$ & $2.9 \times 10^{7}$ & 28,680 &  1522 \\
$^{4}$He   & $N_\mathrm{max} =22$ & $8.6 \times 10^{9}$ & $3.3 \times 10^{5}$ & $2.8 \times 10^5$ \\
$^{12}$C   & $N_\mathrm{shell} =4$ & $5.9 \times 10^{11}$ & $4\times 10^6$ &   $1 \times 10^5$ \\
$^{12}$C   & $N_\mathrm{max} =8$ & $5.9 \times 10^{8}$ & $3 \times 10^6$ & $ 6 \times 10^5$ \\
\hline
\end{tabular}
\end{table}

Furthermore, the fact that the half-Slaters do not have fixed particle number becomes an
 advantage. We do not always need all possible Slater determinants
of a given species, but merely all Slater determinants of a fixed set of
quantum numbers.

For example, consider severe many-body truncations based upon 
the $W$-weighting described in  \ref{weighting}, where one forces 
the sum of weights associated with single-particle states to be less than some value. 
In this case, 
generating the
required Slater determinants, is a nontrivial task. Generating all possible
Slater determinants and eliminating those unneeded turns out to be horribly inefficient.
For example, consider a $W=2$ cut on $^{16}$O. There are only 1245 proton Slater
determinants needed to construct the basis, but if one naively generated all Slater
determinants in the single-particle space there would be 76 million Slater determinants.

The answer, of course, is instead of generating millions or billions of candidate
Slater determinants and throwing away the bulk of them, one creates them recursively:
start by creating one-particle states, then two-particle states, and so on, and by
constraining the additive quantum numbers one can arrive at the required basis. This 
was done, for example, in our first generation code, REDSTICK.  But
even so, this is somewhat wasteful, as one throws away the intermediate basis sets.

The half-Slaters answer this: not only do we construct the half-Slaters recursively
(from the vacuum state we
generate all the needed one-particle half-Slaters, and from the one-particles half-Slaters
we generate all the two-particle half-Slaters, and so on), we actually
\textit{need} all or most of these hSDs in order to create the Slater determinants.
Furthermore, this gives us a route to creating the `jumps.'

Let us explain in more detail.  First, we break the single-particle
space, for example that found in Table \ref{sdstates}. The
single-particle states 1, 3, 5, 7, 8, and 9, which all have $m_j <
0$, are grouped together (as `left' or L states), while the
remaining single particle states with $m_j > 0$ are group as 'right'
or R states. Let's focus first on the left states. We start with the
vacuum: 000000. From the vacuum state we create six one-particle
hSDs: 100000, 010000, 001000, etc. From each of the one-particle
hSDs we create two-particle hSDs; one can do this without
duplication by only adding a particle to the right of all filled
states, that is,
\begin{equation}
001000 \rightarrow 001100, 001010, 001001 \nonumber
\end{equation}
and so on, recursively.  If one is making a cut on the many-body basis using the $W$ weighting,
one simply does not create hSDs that would violate the maximum $W$ allowed.

Almost for free we have the action of creation operators on a half-Slater, that is,
the matrix element
\begin{equation}
\langle \omega^\prime | \hat{a}^\dagger_i | \omega\rangle,
\end{equation}
where $n_{\omega^\prime} = n_\omega + 1$.  We call such a matrix
element a `hop' and it comes out automatically when generating all
the half-Slaters with $n_\omega+1$ particles from the half-Slaters
with $n_\omega$ particles. In practice, we sort the hSDs, already
grouped by the number of particles $n_\omega$, by $M_\omega, \Pi_\omega, W_\omega$; because the
lists are small, the sorting is quick.  Then when computing the
`hop' we know the initial and final quantum numbers and the search
through the list of possible final states is also very quick.  It is
this kind of sorting and searching that is most time-consuming in
occupation-space CI codes, and by breaking down the basis into
half-Slaters, and thus making all lists to be sorted and searched much smaller,  
we speed up the process considerably.

There are typically a small number of these `hops' and they are,
again, organized efficiently by quantum numbers; in fact, given the
quantum numbers associated with $\omega$ and $\omega^\prime$ the
quantum number of the single-particle state $i$ is fixed. Some
illustrative data on hops is found in Table \ref{hops}.

\begin{table}
\caption{Construction of the many-body matrix elements (for two-body
interactions)  via creation-operator `hops.'  For three-body
interactions, the number of jumps and many-body matrix elements will
increase, but the number of hops is fixed. \label{hops}}
\begin{tabular}{|c|c|c|c|c|}
\hline
Nuclide  &  space  & \# nonzero  & \# & \#  \\
         &         &  many-body m.e.s  &  jumps      &  hops \\
\hline
$^{28}$Si  & $sd$ & $2.6 \times 10^6$  & $1.2 \times 10^5$  &  768  \\
$^{52}$Fe  & $pf$ & $8.9 \times 10^{10}$ & $1.2 \times 10^7$ &  15280 \\
$^{56}$Ni  & $pf$ & $1.2 \times 10^{12}$ & $5.6 \times 10^7$ & 20080 \\
$^{4}$He   & $N_\mathrm{shell} =8$&$1.8\times 10^{11}$&$4.3\times 10^7$ & $6 \times 10^4$  \\
$^{4}$He   & $N_\mathrm{max} =22$ & $1.1 \times 10^{12}$  & $5.3 \times 10^9$ & $1.5 \times 10^6$ \\
$^{12}$C   & $N_\mathrm{shell} =3$ & $5 \times 10^{10}$ &  $1 \times 10^7$ & $1.5 \times 10^4$   \\
$^{12}$C   & $N_\mathrm{shell} =4$ & $1.5 \times 10^{15}$ &  $3.3 \times 10^9$ & $1.3 \times 10^6$   \\
$^{12}$C   & $N_\mathrm{max} =8$ & $6.4 \times 10^{11}$ & $3.4\times 10^9$ & $4.2 \times 10^6$ \\
\hline
\end{tabular}
\end{table}

From the hops one can then build one- and two-body jumps
needed for the action of the Hamiltonian. The algorithm for doing so
is slightly involved, but it is quick, because one has dramatically reduced searches. For example, to generate all
the jumps needed for $^{52}$Fe in the $pf$ shell, it takes less than
30 seconds on a standard desktop machine--and each matrix
multiplication takes more than 30 minutes. (For large $W$ cuts, the
efficiencies drop; the number of half-Slaters needed relative to the
final basis dimension is larger, and the time to generate jumps is
also longer.)

In going from REDSTICK, which has a single level of factorization, to BIGSTICK, 
which by exploiting hSDs has two levels of factorization, 
 basis construction speeds up consistently by a factor of 3-4 times, while 
construction of jumps speeds up consistently by a factor of 10, independent of dimensionality.  

  To describe the construction of jumps from hops,  
we have to introduce a few more concepts. 
Along the way, we continue to organize everything by quantum numbers 
$m, \pi, w$. This shortens lists to be searched. 

 Suppose we 
want to build $n$-body jumps, which is the action of $n$ annihilation operators followed by 
$n$ creation operators, e.g. $\hat{a}^\dagger \hat{a}^\dagger \ldots \hat{a} \hat{a}$.
As an intermediate step, we consider $n$-particle 
hops, which  simply chain together $n$ annihilation hops removing $n$ particles.  
Applying an $n$-body annihilation hop to a sector $\Gamma$ creates 
a new, intermediate sector, $Q$, with $n$ fewer particles,  and 
so is not part of the basis,  but is constructed with 
half-Slater determinants already created. 

The key idea, outlined below in pseudocode, is to construct $n$-particle hops from both
 initial and final sectors, $\Gamma^i, \Gamma^f$, to the same intermediate sector $Q$. 
Then, by looping over states $q$ in $Q$, we can work backwards and find
the initial and final states in $\Gamma^i, \Gamma^f$ and construct the jumps.

\medskip

\parbox{11cm}
{
FOR all initial sector $\Gamma^i$

\smallskip

\hfill \parbox{10.5cm}
   {
   Find all allowed $Q$ from subtracting $n$ particles

\smallskip

   FOR all $Q$

\smallskip

\hfill \parbox{10cm}
       {
       Find all $n$-body hops $\Gamma^i \rightarrow Q$

\smallskip

       FOR all final sectors $\Gamma^f$
 
\smallskip 

\hfill \parbox{9.5cm}
            {     
            Find all $n$-body hops $\Gamma^f \rightarrow Q$

\smallskip

           FOR all intermediate states $q \in Q$

\smallskip

\hfill \parbox{9cm}
                 {
                  Find all initial $\mu \in \Gamma^i$, such that $\mu \rightarrow q$

\smallskip
  
                  Find all all final $\mu^\prime \in \Gamma^f$, such that $\mu^\prime \rightarrow q$

\smallskip 

                 Create $n$-body jump(s) $\mu \rightarrow \mu^\prime$

                   }
\smallskip

             END FOR
            }

\smallskip

        END FOR
       }

\smallskip
  
    END FOR
   }

\smallskip

END FOR
}

\medskip

Because the hops carry information about the fermion creation and annihilation operators, that 
information is automatically inherited by jumps, allowing one to find the appropriate 
matrix element.

Parceling the basis and the interaction also allows for rather fine
control for parallel work, as discussed in detail in the next
section.

\section{Parallelization and load balancing}

\label{parallel}

All the CI codes described in this paper follow Whitehead's
pioneering use of the Lanczos algorithm to efficiently find
low-lying eigenstates\cite{Lanczos}. (Another popular algorithm is Davidson-Liu
and its variants \cite{Dav93,LSAS01}, designed primarily for
diagonal-dominant matrices, a context appropriate for atomic physics
but not nuclear.) The central operation in the Lanczos algorithm is
a matrix-vector multiplication (matvec). Since for an $M$-scheme
basis the Hamiltonian many-body matrix is very sparse, the key to
efficient parallelization and load balancing of the matvec operation
is to distribute as evenly as possible the nonzero matrix elements
across the compute nodes.

\begin{figure}
\includegraphics [width = 7.5cm,clip]{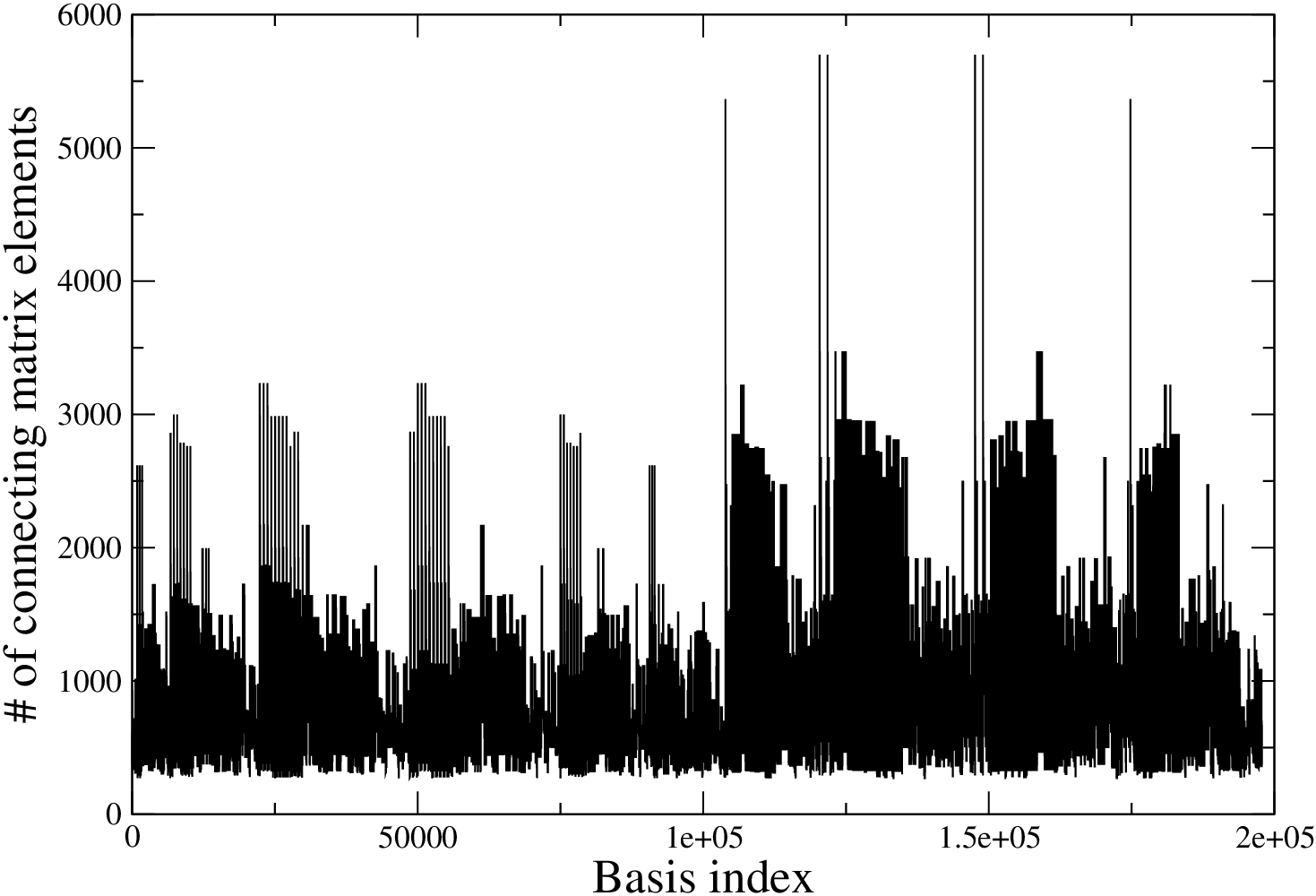}
\caption{\label{medistro} For $^{6}$Li in $N_\mathrm{max} = 6$ space
(dimension = 197,822), number of matrix elements connecting to each
basis state.
}
\end{figure}

\begin{figure}
\includegraphics [width = 7.5cm,clip]{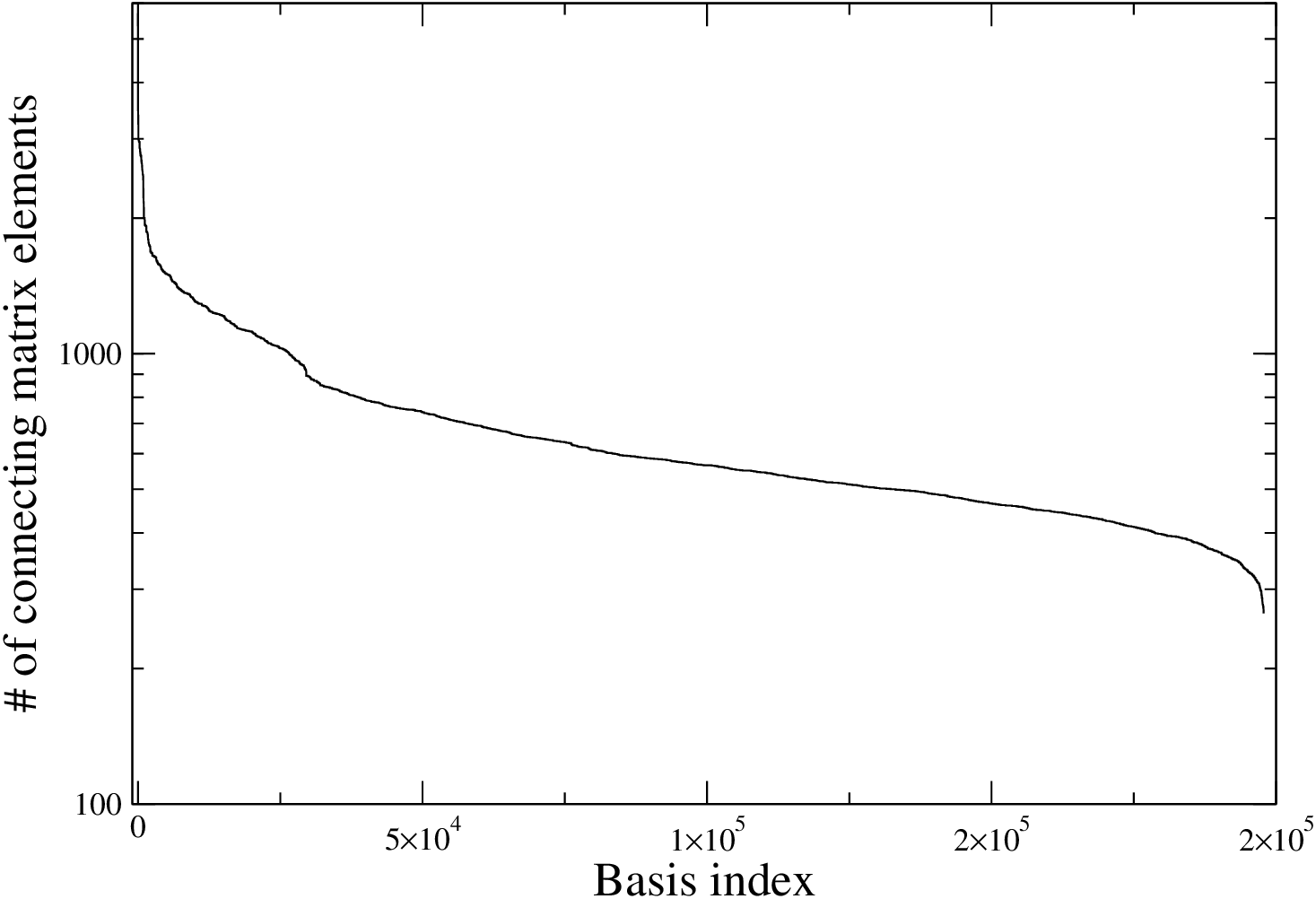}
\caption{\label{medistrosort} The same as Fig.~\ref{medistro},
but sorted on the number of connecting matrix elements..
}
\end{figure}

In our first-generation code, REDSTICK, we distributed the matvec
operations by initial basis state.  Each basis state, however, has
vastly different numbers of operations connection to it.
Fig.~\ref{medistro} illustrates this for the case of $^{6}$Li in an
$N_\mathrm{max} = 6$ space (two-body interactions only);
Fig.~\ref{medistrosort} shows on a log graph the same information
but sorted. Thus, naively distributing work by basis state lead to
unbalanced workloads. Sorting as in Fig.~\ref{medistrosort} will not
help to balance workloads as ``nearby'' initial states connect to
very different final states.

The partial grouping can be understood as related to organizing the
basis into sectors labeled by quantum numbers of the proton Slater
determinants.  When storing matrix elements on core, one can average
this out through a randomizing, `round-robin' algorithm
\cite{LanczosStore}. Such a method will not work when using quantum
numbers to factorize the Hamiltonian, however.

While investigating the distribution of the matvec operations, we
realized that factorization itself gives us the key to
parallelization. As explained in section III, the Hamiltonian is a
sum of one- and two-body terms. Since it is easy to work with the
one-body terms, here we focus on the parallelization of the
application of the two-body terms, i.e., $\hat{H}_{xx}$,
$\hat{H}_{yy}$, and $\hat{H}_{xy}$. The action of these operators is
organized in one- and two-body jumps. The key to efficient
parallelization of BIGSTICK is to distribute the jumps across
parallel cores in such a way that each compute core performs
(approximately) equal number of operations as described below. (A
`core' is either a MPI process, in a pure MPI distributed memory
programming model, or a thread, in an OpenMP shared memory programming model, or a 
hybrid MPI+OpenMP programming model.)

Rather than counting up operations individually to/from each state,
we break up the Hamiltonian into blocks based upon quantum numbers,
that is, we organize Hamiltonian operations by the initial and final
sectors of the basis (where a `sector' of the basis is labeled by
the quantum number $M_x, \Pi_x, W_x$ of the $x$-species). In this
way, we can compute the number of operations without actually
counting them individually. In the example discussed in Section
III.C, we know from 2677 proton two-body jumps and 109 conjugate
neutron Slater determinants, that we get $2677 \times 109 =291,793$
operations.
It is these `blocks' of operations, based upon quantum numbers, that
we break up and distribute across compute cores, with fairly modest
bookkeeping.

For instance, continuing the same example, suppose we find we want
approximately 100,000 operations per compute core. For the block of
operations in our example, we can have proton two-body jumps number
1 through 917 on the first compute core, jumps 918 through 1834 on
the second compute core, and 1835 through 2677 on the third compute
core.  Each core loops over all 109 conjugate neutron Slater
determinants.

\begin{figure}
\includegraphics [width = 7.5cm]{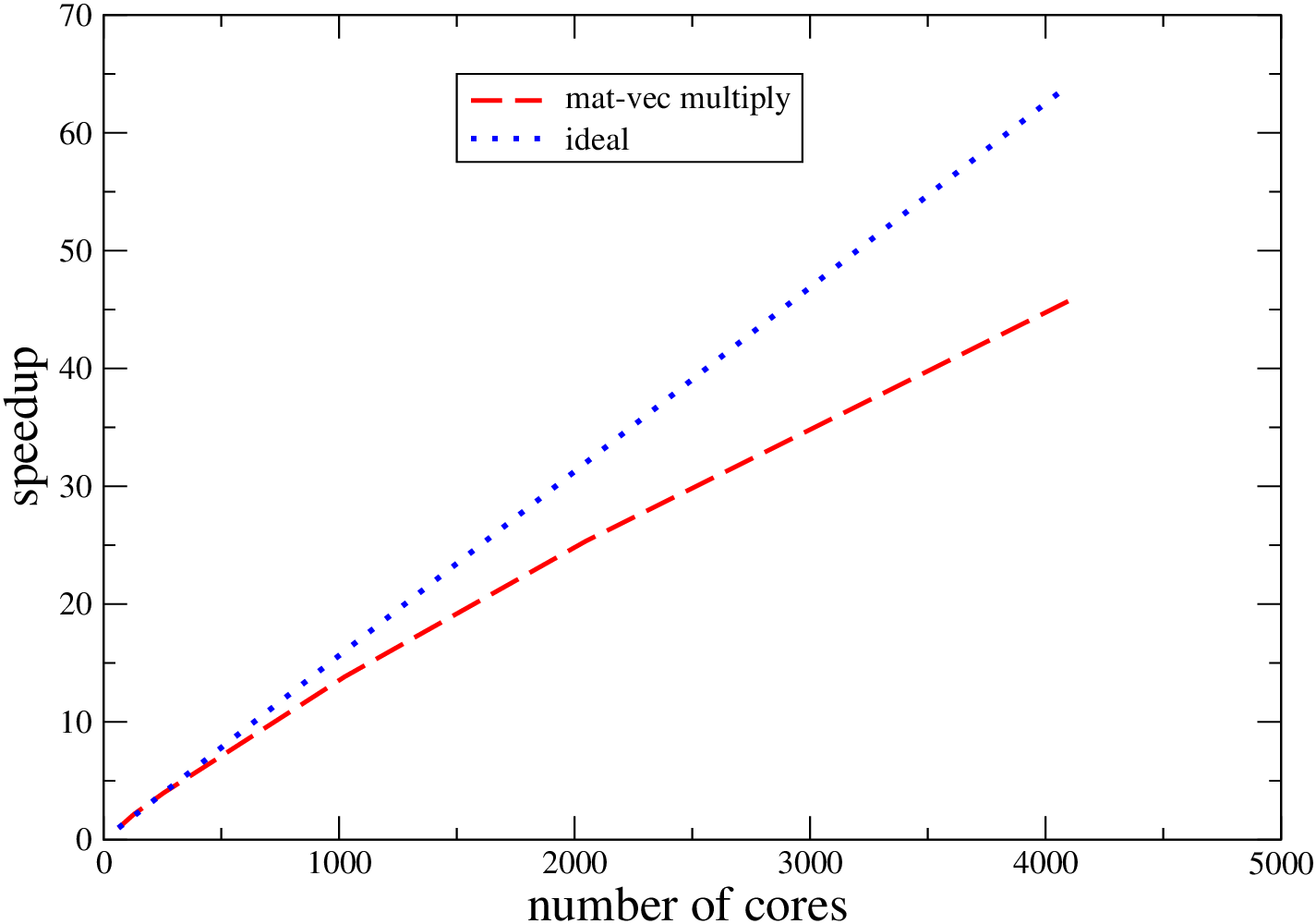}
\caption{\label{speedup_fig} (Color online) Relative speed-up of the matvec
operation for $^{50}$Mn in the $pf$-shell, with an $M$-scheme dimension of
$1.8 \times 10^7$ with three-body interactions. Solid (black) line corresponds to the ideal speed-up and
broken (red) line to the actual speed-up. Data from Jaguar at Oak Ridge National Laboratory in September 2012.}
\end{figure}

We find this distribution of the matrix-vector multiplication scales
very well. Fig.~\ref{speedup_fig} demonstrates scaling from 64 to
4096 cores for the case of $^{50}$Mn in the $pf$ shell (basis
dimension of 18 million) with a three-body force; the data was taken
on the Jaguar supercomputer at Oak Ridge National Laboratory in
September 2012. We have done a number of similar studies; for
example, with an earlier version of BIGSTICK we found similar
scaling of the matvec operation from 500 to 10000 cores on the
Franklin machine at the National Energy Research Computing Center
(NERSC) for  $^{52}$Fe in the $pf$ shell (dimension 110 million)
with a two-body interaction.


At this point the major computation bottlenecks are communication and, if
one writes the Lanczos vectors to disk, I/O (alternately, one can
store the Lanczos vectors also across many compute nodes\cite{LanczosStore}).  All parallel CI codes
face these same barriers.

\section{Summary and Conclusion}

Configuration-interaction calculations of the many-fermion problem
are straightforward: one  diagonalizes the Hamiltonian in a
many-body basis. If that basis is made of Slater determinants with
fixed $J_z$, the resulting matrix is very sparse; despite the
sparsity, however, the number of nonzero matrix elements quickly
becomes overwhelming (cf. Table \ref{nuclear_sparsity}).
Investigation of the nonzero matrix elements reveals a high degree of
redundancy, that is, the same numerical value of the  matrix
element, up to a phase, over and over again (cf. Table
\ref{redundancy}).

Both the sparsity and the redundancy can be understood as the
consequence of having a two- or three-body interaction embedded in a
many-body space, with a consequence of many spectator particles.
Nonetheless, while the sparsity reduces the number of many-body
matrix elements, simply determining the nonzero matrix
elements--often of the order of just one in  a million--is a
nontrivial problem.

That very sparsity, however, can be turned to an advantage. In this
paper, we discussed how, if one has two species of particles and if
there are abelian quantum numbers (which simply means the quantum
numbers are additive or multiplicative),  one can further compactify
the problem via factorization. Factorization yields a memory savings
of one to almost  three orders of magnitude, as shown in Tables
\ref{jumps},\ref{threebody}; furthermore, it can make planning of
the parallel distribution of work straightforward, as one can
calculate the number of operations without explicitly constructing
them. While factorization has been used in nuclear physics codes for
well over a decade, recent work taking factorization an additional
step leads to additional efficiencies. There is a cost to all this,
of course, in more complicated internal bookkeeping.  It is worth
pursuing, however, as it will allow us to tackle significantly
larger problems, than a straight matrix storage scheme, on the same
computing platform.

The U.S.~Department of Energy supported this investigation through
contracts DE-FG02-96ER40985 and DE-FC02-09ER41587, and through
subcontract B576152 by Lawrence Livermore National Laboratory under
contract DE-AC52-07NA27344. We would like to thank H.~Nam, J. Vary, and
P. Maris for helpful conversations over the years regarding the development 
of CI codes.

\appendix

\section{Quantum numbers}

\label{quantum_numbers}

In this section, we briefly introduce the concept of quantum numbers for non-physicists
(e.g., computer scientists and applied mathematicians).

Quantum numbers label irreducible representations of a group. In quantum mechanics,
the labeling is done using eigenvalues of commuting operators \cite{GroupRep}; often
these eigenvalues in turn correspond to physically conserved quantities, especially if
the group is continuous.

In classical mechanics, Noether's theorem states that
if a Hamiltonian is invariant under an operation, there will be
a corresponding conserved quantity. Invariance under spatial translation leads to
conservation of linear momentum, while invariance under spatial
rotation leads to conservation of angular momentum.

In quantum mechanics, any observable quantity that can be measured is represented by a Hermitian,
linear operator $\hat{\cal O}$; the allowed values that can be measured are the
eigenvalues of the operator,
\begin{equation}
\hat{\cal O} | \Psi \rangle = \lambda | \Psi \rangle.
\end{equation}
Invariance in quantum mechanics is given by commutation: if $[ \hat{H}, \hat{\cal O} ] = 0$,
then one can have simultaneous eigenstates:
\begin{equation}
\hat{H} | E,\lambda \rangle = E | E,\lambda \rangle, \,\,\,\,
\hat{\cal O} | E,\lambda \rangle = \lambda | E,\lambda \rangle.
\end{equation}
In this case, $\lambda$ corresponds to a conserved quantity: this value  does not change
under the evolution of the wavefunction through the time-dependent Sch\"odinger equation.
For many important conserved quantities, such as total angular momentum
\begin{equation}
\hat{J}^2 | \Psi \rangle = j(j+1) \hbar^2 | \Psi \rangle
\end{equation}
and the $z$-component of angular momentum
\begin{equation}
\hat{J}_z | \Psi \rangle = m \hbar | \Psi \rangle
\end{equation}
the eigenvalues can be expressed in simple numbers: here $j, m$ are either integers
or half-integers. (The convention in this paper is we use lower case letters $j,m$, etc., for
single-particle quantum numbers, and capital letters $J,M$, etc., for quantum numbers of many-body
systems.)  This is the origin of the term `quantum numbers,' which in physics refers
to eigenvalues that are exactly or approximately conserved.

Conserved quantities
lead to \textit{selection rules}, which mean that certain matrix elements must be zero. For example,
with a rotationally invariant Hamiltonian, total $M$ is conserved, and hence
there can be no matrix elements between basis states of different total $M$.
Physically, $M$ is related to orientation, which for a system governed by a rotationally
invariant Hamiltonian cannot change spontaneously.

For a composite system, one needs group theory to combine quantum numbers.
Important to us is the fact that some quantum numbers can be combined simply:
$J_z$ is added, and parity is multiplied. (Technically, this is because they
are represented by abelian groups.)  Other quantum numbers,
such as total angular momentum, associated with non-abelian groups require more sophisticated tensor algebra. It is the
simplicity of additive and/or multiplicative quantum numbers that we exploit here. It
is also why the $M$-scheme basis, constructed from a simply additive (abelian) quantum
number is numerically easy, and why the $J$-scheme basis, while more compact,
is also more challenging, constructed from  non-abelian quantum numbers.

\section{Model space for nuclei}
\label{nuclear_models}

For the nonexpert, we describe here several typical model spaces for
CI calculations of low-energy nuclear structure.  When doing \textit{ab initio}
or \textit{no-core shell-model} calculations, one uses harmonic oscillator single-particle
orbits $0s,0p, 1s,0d$, and so on, where the first number is $n$, the number of nodes
in the radial wavefunction. Because spin-orbit splitting is large in nuclei,
one also appends total $j$: $0s_{1/2}, 0p_{1/2}, 0p_{3/2}$, etc.. These orbits are
grouped together into \textit{shells} (sometimes called \textit{major shells})
by the principal quantum number $N= 2n+1$. Hence $0s_{1/2}$ has $N=0$,
$0p_{1/2}, 0p_{3/2}$ have $N=1$, $1s_{1/2}, 0d_{3/2}, 0d_{5/2}$ have $N=2$ and so on.
Below, in  \ref{weighting} we describe various truncation schemes, the
most important for these no-core calculations are either truncating solely on
the number of shells, so that $N_\mathrm{shell}$ = 3 include the $N= 0,1,2$ shells;
or, the $N_\mathrm{max}$ or $N\hbar \Omega$ truncation scheme.

The latter is
described in more detail in  \ref{weighting}, but briefly: to each many-body
state one assigns an energy which is the sum of the non-interacting harmonic
oscillator energies, leaving off the zero-point energy of $\frac{3}{2} \hbar \Omega$ per particle.
In that case, one includes only states up to a certain cutoff in
this energy above the lowest possible energy.  For example, consider
$^{4}$He. If all four nucleons are in the $0s_{1/2}$ orbit, the non-interacting
energy is $0\hbar\Omega$.
(Here $\hbar\Omega$ simply provides the energy scale but the value of $\Omega$ does
not affect the truncation.) If one excites 1 particle from the $0s_{1/2}$ orbit to,
say, the $2s_{1/2}$ or the $1d_{5/2}$ or the $0g_{9/2}$ orbits (all in the
$N=4$ or $2s$-$1d$-$0g$ major shell), then the state has a non-interacting energy of $4\hbar\Omega$.
One can achieve the same value by exciting two particles up to the $N=2$ major
shell, or all four up to the $N=1$ or $0p$ shell, or 1 particle to $N=1$ and another up to
$N=3$.  We call the set of all states with non-interacting energy equal to or less
than $4\hbar\Omega$ the $N_\mathrm{max} = 4$ space. It gets more complicated for
more particles. For example, for $^{12}$C the lowest non-interacting energy
is in fact $8\hbar\Omega$ (4 particles in the $N=0$ or $0s$ shell and the rest in the $N=1$
or $0p$ shell). An $N_\mathrm{max}=4$ \textit{excitation} includes exciting one
particle from the $0s$ up to the $2s$-$1d$-$0g$ shell, \textit{or} from the
$0p$ ($N=1$) up to the $2p$-$1f$-$0h$ ($N=5$), \textit{or} two particles from the $0p$ up 
to the $1p$-$0f$, or 4 particles from the $0p$ up to the $1s$-$0d$, etc.

We also use as examples two valence model spaces.  The first is the
$1s$-$0d$ shell, also known more simply as the $sd$-shell. The valence space we choose to
be the $1s_{1/2}$, $0d_{3/2}$ and $0d_{5/2}$ orbits, also known
as the $sd$ or $1s0d$ space, the quantum
numbers of which are given in Table \ref{sdstates}, along with an
inert, frozen $^{16}$O core, that is,  $(0s_{1/2})^4 (0p_{1/2})^4 (0p_{3/2})^8$
filled configuration. (Because of the strong spin-orbit force in nuclei, even in
the simplest approximations one must consider not only $l$ but also
$j$ of each single-particle state.)   In addition, we consider the
$pf$ shell, which assumes an inert $^{40}$Ca core and with an active valence
space the $1p_{1/2}, 1p_{3/2}, 0f_{3/2}$ and $0f_{5/2}$ orbits.
Throughout the
rest of this paper we will give examples built  upon these spaces.

\section{Examples from atomic physics}

\label{atomic_examples}

Our examples in the text were primarily taken from low-energy nuclear
physics. In order to provide a different context, we present here some
examples from atomic physics,  electrons around a single atom and
cold spin-$1/2$ atoms in a harmonic trap. In both cases we treat ``spin-up''
and ``spin-down'' particles as separate species (in place of protons and
neutrons); thus the orbits are labeled by $l$ rather than $j$.  All fermionic
properties are preserved, however.

To begin, we demonstrate how to construct the many-body basis for electrons
around an atom, using the small model space in Table \ref{3SPDstates}.
We assume a  frozen Ne core, that is,
frozen $(1S)^2 (2S)^2 (2P)^6$, and have valence $3S$, $3P$, and $3D$ states.
Consider specifically the case of neutral phosphorus, with five valence electrons.
We take three `up' electrons and two 'down' electrons; if the force lacks a spin-orbit
component, then both total $\vec{L}$ and total $\vec{S}$ will be good quantum numbers.

\begin{table}
\caption{Ordered list of single-particle states in the $N=3$-shell for atomic electrons.
\label{3SPDstates}}
\begin{tabular}{|c|c|r|}
\hline
State & $l$ & $m_l$ \\
\hline
1 & 0 & 0  \\
2 & 1 & -1 \\
3 & 1 & 0 \\
4 & 1 & 1 \\
5 & 2 & -2 \\
6 & 2 & -1 \\
7 & 2 & 0 \\
8 & 2 & 1 \\
9 & 2 & 2 \\
\hline
\end{tabular}
\end{table}

We can construct Slater determinants as we did for nuclear examples. So,
using the number from Table \ref{3SPDstates}, the  Slater determinant for
three spin-up particles
$\hat{a}_2^\dagger \hat{a}_4^\dagger \hat{a}^\dagger_9 |0 \rangle$,
which can be represented in bit form as 010100001, has total $M = L_z = 2$ and
parity $+$. Because we assume the same orbits for spin-down, Slater determinants
for spin-down electrons are similar.

When we construct all the states with three spin-up, two spin-down, total $L_z = 0$
and total parity = $+$, as shown in Table \ref{phosphorus}: there are 252 such states.
Note that we use all $\left ( \begin{array}{c} 9 \\ 2 \end{array} \right) = 36$
of the spin-down Slater determinants, but we are missing two of the  $\left ( \begin{array}{c} 9 \\ 3 \end{array} \right) = 82$; those are the states with
$M_\uparrow = \pm 4$.

\begin{table}
\caption{Decomposition of the $M$-scheme basis for 3 spin-up and 2
spin-down electrons, using the valence space of Table \ref{3SPDstates} and
assuming a frozen Ne core (hence phosphorus), with  total $L_z = M_\uparrow + M_\downarrow 0$ and total parity $\pi = \pi_\uparrow \times \pi_\downarrow$. Here ``$\uparrow$ SD'' =  Slater determinants composed of spin-up electrons and ``$\downarrow$SD'' =
 Slater determinants composed of spin-down electrons, while ``combined'' refers to the
combined  many-body basis states. \label{phosphorus}}
\begin{tabular}{|r|r|r|r|r|}
\hline
$M_\uparrow^{\pi_\uparrow}$ & \# $\uparrow$SDs  & $M_\downarrow^{\pi_\downarrow}$ & \# $\downarrow$SDs & \# combined \\
\hline
$3^+$ & 3 & -3$^+$ & 1 & 3 \\
$3^-$ & 3 & -3$^-$ & 1 & 3 \\
$2^+$ & 4 & -2$^+$ & 2 & 8 \\
$2^-$ & 6 & -2$^-$ & 2 & 12 \\
$1^+$ & 8 & -1$^+$ & 4 & 32 \\
$1^-$ & 8 & -1$^-$ & 4 & 32\\
$0^+$ & 8& 0$^+$ & 4 & 32 \\
$0^-$ & 10 & 0$^-$ & 4 & 40 \\
-$1^+$ & 8 & 1$^+$ & 4 &32 \\
$-1^-$ & 8 & 1$^-$ & 4 &32 \\
$-2^+$ & 4 & 2$^+$ & 2 & 8 \\
$-2^-$ & 6 & 2$^-$ & 2 & 12 \\
$-3^+$ & 3 & 3$^+$ & 1 & 3 \\
$-3^-$ & 3 & 3$^-$ & 1 & 3 \\
\hline
Total & 82 & & 36 & 252 \\
\hline
\end{tabular}
\end{table}

We also consider the valence space composed of $N=3,4,5$ hydrogen-like orbitals, not shown.  The details of the orbitals, i.e. the radial dependence, 
is not crucial to our point here, which  is to 
illustrate comparative memory requirements.   

Again, taking phosphorus with $M^\pi = 0^+$ basis states, in the $N=3$ valence space we can construct 242 basis states from 118 Slater determinants, 
as illustration in Table \ref{phosphorus}, which in turn can be constructed from 89 hSDs. For the equivalent basis in the $N=3,4,5$ space, 
the basis dimension is 1.5 million, constructed from 20,000 Slater determinants and 6800 hSDs.

Looking at the Hamiltonian matrices, for the $N=3$ space the sparsity is about $20\%$ and with a redundancy of 18, that is every unique matrix element 
is reused on average 18 times in the many-body Hamiltonian matrix. For the $N=3,4,5$ space, the sparsity is $0.2 \%$ and the redundancy is 3400. 

The Hamiltonian matrix for the $N=3$ space requires 12,500 operations, constructed from 1847 jumps, built from 153 hops; the memory requirement
for the nonzero many-body matrix elements is  0.5 Mb, while storage of the jumps requires only 0.02 Mb. The Hamiltonian matrix for the $N=3,4,5$ space, 
1.5 billion operations, requires about 6 Gb of memory, while the corresponding 3.6 million jumps  require only 47 Mb, which in turn are built from 
only 19,000 hops.

\section{Further truncation of the many-body basis}
\label{weighting}

Often one wants to further truncate the many-body basis, either on the
basis of physics or on simple computational efficiency. We will discuss
two common truncation schemes, and then introduce a relatively simple and flexible
generalization that encompasses these cases. (Nonetheless the
following is rather technical and casual readers can skim the following section
without loss of comprehension for the rest of the paper.)

In all our considerations we truncate the many-body space based upon
the single particle space, that is, upon single-particle quantum
numbers. One could truncate based upon many-body quantum numbers,
but that is beyond the scope of this paper and of our algorithms.

\begin{figure}
\includegraphics [width = 7.5cm]{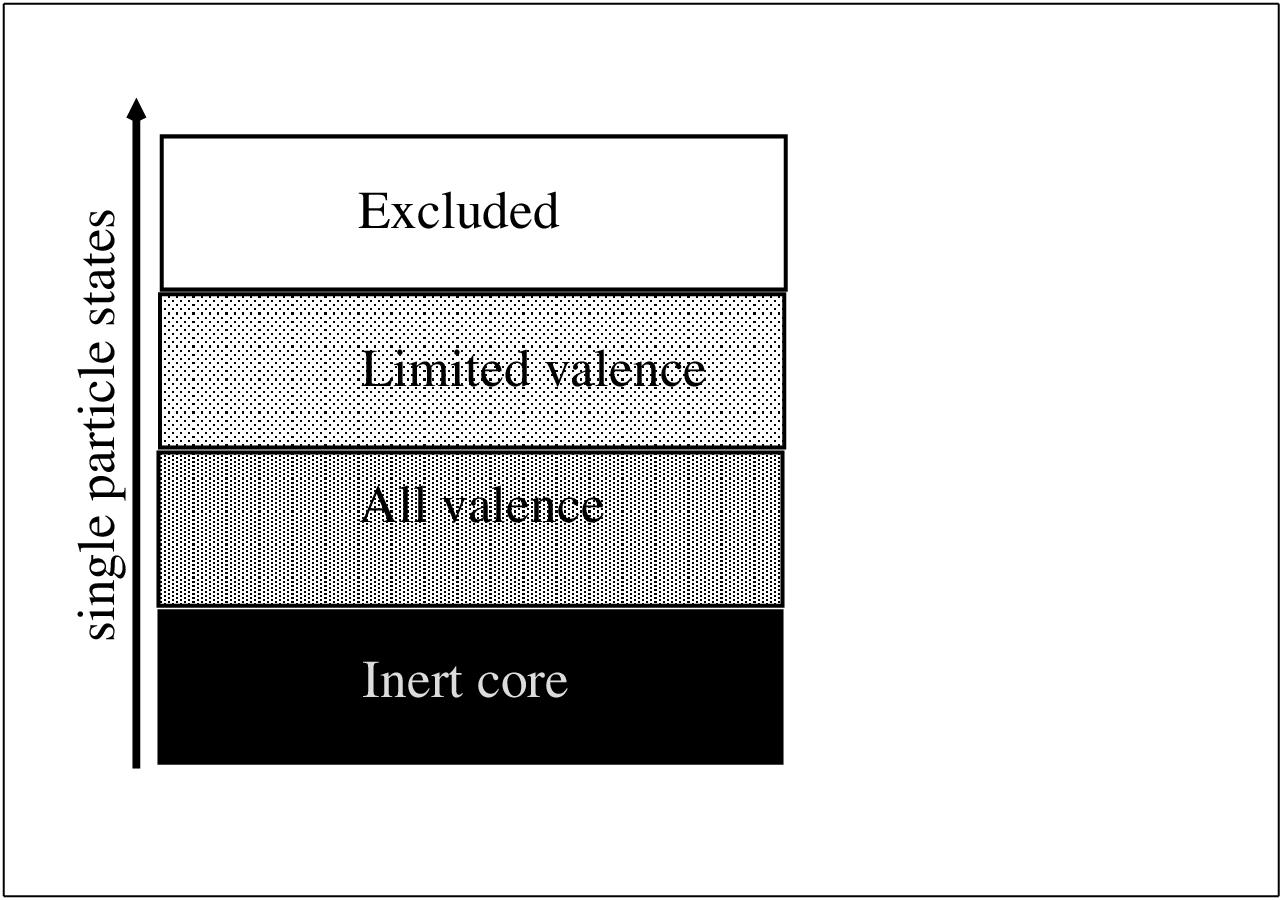}
\caption{\label{valence} Segregation of single-particle space. 'Inert core'
has all states filled. `Excluded' disallows any occupied states.  `All valence' can 
have states up to the number of valence particles filled, while `Limited valence' 
can only have fewer states filled (e.g. one, two, three...). See  text for discussion.
}
\end{figure}

The first kind of truncation is sometimes called a particle-hole
truncation in nuclear physics; in atomic physics (and occasionally
in nuclear physics), one uses the notation `singles,' `doubles,'
`triples,' etc. To understand this truncation scheme, begin by
considering a space of single-particle state, illustrated in Figure
\ref{valence}.  The single-particle space can be partitioned into
four parts. In the first part, labeled `inert core', the states are
all filled and remain filled. In the fourth and final part, labeled
`excluded,' no particles are allowed. Both the core and excluded
parts of the single-particle space need not be considered
explicitly, only implicitly. In some cases there is no core.

More important are the second and third sections, labeled `all
valence' and `limited valence', respectively.  The total number of
particles in these combined sections is fixed at $N_v$, and this is
the valence or active space.

The difference between the `limited valence' and the `all valence'
spaces is that only some maximal number $N_l < N_v$ of particles are
allowed in the 'limited valence' space. So, for example, suppose we
have four valence particles, but only allow at most two particles
into the 'limited valence' space. In this case the `all valence'
might contain four, three, or two particles, while the 'limited
valence' space might have zero, one, or two particles. In more
standard language, $N_l = 1$ is called `one-particle, one-hole' or
`singles',  while $N_l = 2$ is called `two-particle, two-hole' or
'doubles', and so on. There are no other restrictions aside from
global restrictions on quantum numbers such as parity and $M$.

In nuclear structure physics, where center-of-mass considerations
weigh heavily, one sometimes invoke a weighted refinement of this
scheme. For all but the lightest systems, one must work in the
laboratory frame, that is, the wavefunction is a function of
laboratory coordinates, $\Psi = \Psi(r_1, r_2, r_3,\ldots)$. It is
only the relative degrees of freedom that are relevant, however, so
ideally one would like to be able to factorize this into relative
and center-of-mass motion:
\begin{equation}
\Psi(r_1,r_2, r_3,\ldots) = \Psi_\mathrm{rel}(\vec{r}_1 -\vec{r}_2,
\vec{r}_1 - \vec{r}_3, \ldots ) \times \Psi_\mathrm{CM} (\vec{R}_\mathrm{CM} )
\end{equation}
(note that we have only sketched this factorization).
In a harmonic oscillator basis and with a translationally
invariant interaction, one can achieve this factorization exactly, \textit{if}
the many-body basis is truncated as follows \cite{Pal67,PP68,GL74}:

$\bullet$ In the non-interacting harmonic oscillator, each
single-particle state has an energy $e_i = \hbar \Omega (N_i +
3/2)$. Here $N_i$ is the principal quantum number, which is 0 for
the $0s$ shell, 1 for the $0p$ shell, 2 for the $1s$-$0d$ shell, and
so on. The frequency $\Omega$ of the harmonic oscillator is a
parameter but its numerical value plays no role in the basis truncation.

$\bullet$ We can then assign to each many-body state a
non-interacting energy $E_{NI} = \sum_i e_i$, the sum of the
individual non-interacting energies of each particle.   There will
be some minimum $E_\mathrm{min}$ and all subsequent non-interacting
energies will come in steps of $\hbar \Omega$--in fact for states of
the same parity, in steps of $2\hbar \Omega$.

$\bullet$ Now choose some $N_\mathrm{max}$, and allow only states with non-interacting energy
$E_{NI} \leq E_\mathrm{min} + N_\mathrm{max} \hbar \Omega$. In practice, restricting
states to the same parity means that the `normal' parity will have
$E_{NI} = E_\mathrm{min}$,  $E_\mathrm{min}+ 2 \hbar \Omega$,
$E_\mathrm{min}+ 4 \hbar \Omega, \ldots$, $E_\mathrm{min}+  N_\mathrm{max} \hbar \Omega$, while `abnormal' parity will have
$E_{NI} = E_\mathrm{min}+  \hbar \Omega$,
$E_\mathrm{min}+ 3 \hbar \Omega, \ldots$,
$E_\mathrm{min}+  N_\mathrm{max} \hbar \Omega$.

This is sometimes call the $N\hbar\Omega$ truncation, or simply the
energy truncation. It is more complicated than the previous
`particle-hole' truncation. We identify with each principal quantum
number $N_i$ a major shell; for a $4\hbar\Omega$ we can excite four
particles each up one shell, one particle up four shells, two
particles each up two shells, one particle up one shell and another
up three shells, and so on. While complicated, such a truncation
allows us to guarantee the center-of-mass wavefunction is a simple
Gaussian.

Both truncation schemes can be described by introducing an
additional additive quantum number, which we call the weighting $w$
for single-particle states and $W$ for many-body states.  Assign to
each single-particle state $\phi_i$ a non-negative integer $w_i$;
for example, this might be the principal quantum number for the
spherical harmonic oscillator. We assume that states of a given
single-particle level (that is, labeled by unique $j_i, \pi_i,
\alpha_i$ but having distinct $m_i$) share the same $w_i$.  The
weighting is additive, so that, like $M$, the $W$ for a given
many-body state is simply the sum of the $w_i$s of the occupied
states.

Now, the truncation is simply defined by: allow all states with $W
\leq W_\mathrm{max}$. (Usually $W_\mathrm{max}$ is defined, as in
the $N\hbar\Omega$ truncation, relative to some $W_\mathrm{min}$.)
To regain the simple `particle-hole' truncation, we assign $w = 0$
to the `all valence' single-particle states and $w = 1$ to the
`limited valence' states, and set $W_\mathrm{max} = N_l$. One can
devise, however, more complicated weightings.  Because of the inequality 
and not a strict equality, $W$ is not a `good quantum number.'  However 
we can use nearly the same machinery to implement the inequality 
as the equality. 

This weighted truncation can be and has
been implemented into our CI code.

\end{document}